\documentclass[iicol,sn-basic]{sn-jnl}

\usepackage{graphicx}
\usepackage{multirow}
\usepackage{amsmath,amssymb,amsfonts}
\usepackage{amsthm}
\usepackage{mathrsfs}
\usepackage[title]{appendix}
\usepackage{xcolor}
\usepackage{textcomp}
\usepackage{manyfoot}
\usepackage{booktabs}
\usepackage{algorithm}
\usepackage{algorithmicx}
\usepackage{algpseudocode}
\usepackage{listings}
\usepackage{float}
\floatstyle{plaintop}
\usepackage{tabularx}
\usepackage{booktabs}
\usepackage{siunitx}
\usepackage{array}
\usepackage{algpseudocode}
\usepackage{mathtools}

\theoremstyle{thmstyleone}%
\theoremstyle{thmstyletwo}%

\theoremstyle{thmstylethree}%

\begin{document}

\title{Variational phase-field modeling of fracture and fatigue in shape memory alloys: a one-dimensional study}

\author[1]{\fnm{Alma} \sur{Brambilla}}\email{alma.brambilla@polimi.it}

\author*[2]{\fnm{Laura} \sur{De Lorenzis}}\email{ldelorenzis@ethz.ch}

\author[1]{\fnm{Lorenza} \sur{Petrini}}\email{lorenza.petrini@polimi.it}

\affil[1]{\orgdiv{Department of Civil and Environmental Engineering}, \orgname{Politecnico di Milano}, \orgaddress{\street{Piazza Leonardo da Vinci 32}, \city{Milan}, \postcode{20133}, \country{Italy}}}

\affil[2]{\orgdiv{Department of Mechanical and Process Engineering}, \orgname{ETH Zürich}, \orgaddress{\street{Tannenstrasse 3}, \city{Zürich}, \postcode{8092}, \country{Switzerland}}}

\abstract{We propose a novel variational phase-field model for fracture and fatigue in pseudoelastic shape memory alloys (SMAs). The model, developed in a one-dimensional setting, builds upon the Auricchio–Petrini constitutive formulation for SMAs and couples damage evolution with phase transformation. We study analytically and numerically the homogeneous and localization responses of a bar under both monotonic and cyclic loading, and we investigate various macroscopic behaviors by tuning the constitutive parameters. 
A key feature of the model is the introduction of a transformation strain limit, beyond which the material is fully martensitic and behaves elastically. This leads to a distinctive behavior in which the region of localized damage widens, yielding a delay of fracture.
The capability of the model to predict the fatigue performance is assessed by simulating the uniaxial response of Ni-Ti multi-wire samples under different loading conditions. The results show that the model discriminates between safe and critical loading scenarios, capturing the experimental trend of increased fatigue resistance with higher mean strain at a fixed strain amplitude. Ongoing efforts are aimed at further evaluating its reliability for quantitative fatigue life prediction.}


\maketitle

\section{Introduction}
\begin{figure*}[t]
    \centering
    \includegraphics[width=0.8\textwidth]{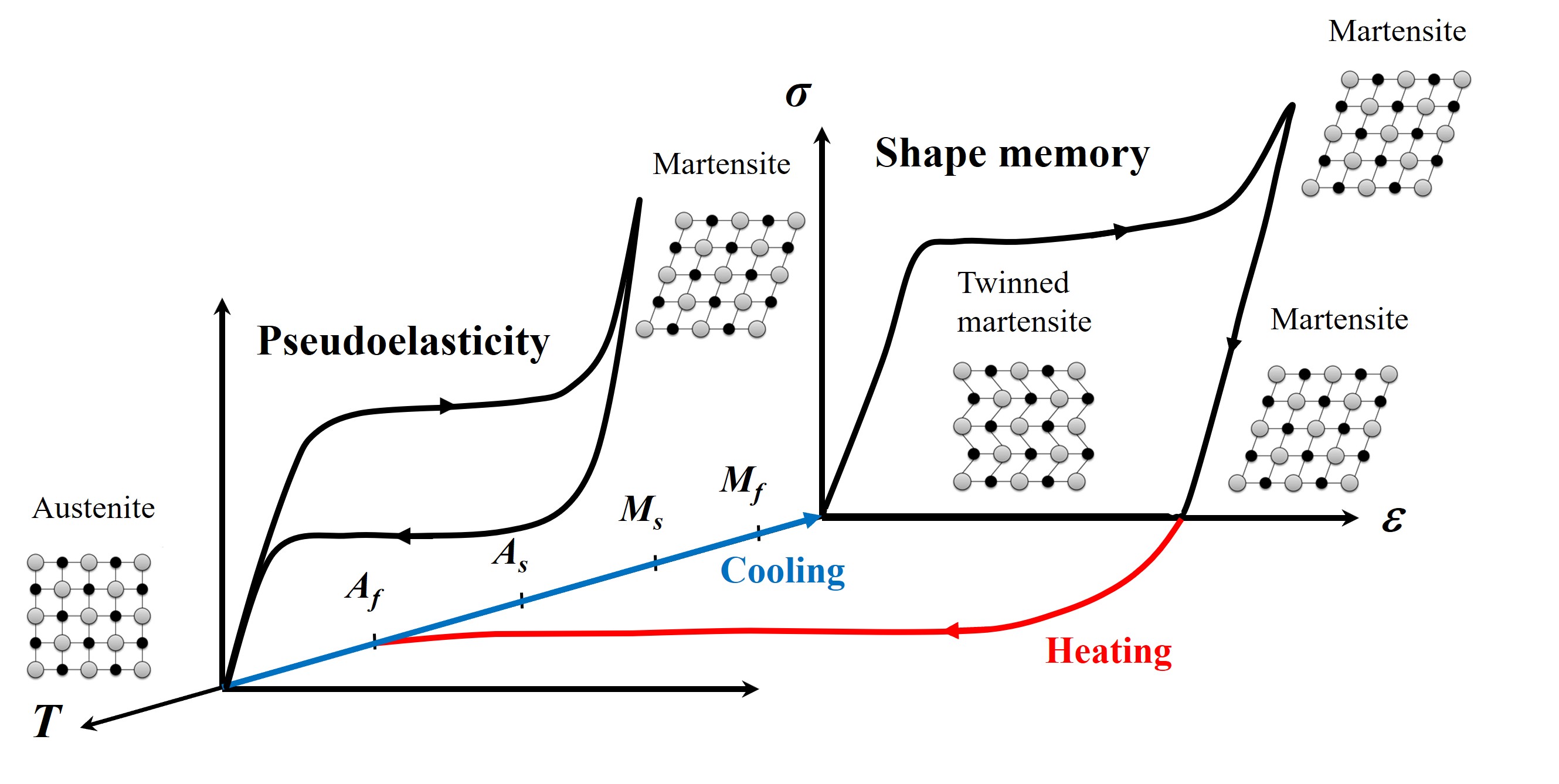}
    \caption{Characteristic thermomechanical behavior of SMAs: (i) pseudoelasticity above the austenite finish temperature $A_f$ allowing large stress-induced strain recovery, and (ii) shape memory effect allowing the recovery of the original shape after deformation in the martensitic phase below the martensite finish temperature $M_f$.}
    \label{fig1}
\end{figure*}

Shape memory alloys (SMAs) are a class of intermetallic materials which display non-conventional thermomechanical properties arising from a reversible diffusionless thermoelastic transformation between the austenite and the martensite phase \citep{Lagoudas2008}. This transformation can be attained under both mechanical or thermal stimuli, resulting in two characteristic behaviors at the macroscopic level: the shape memory effect and pseudoelasticity, schematically depicted in Figure \ref{fig1}. Each alloy is characterized by specific transformation temperatures $M_s$, $M_f$, $A_s$, and $A_f$, representing the martensite start and finish temperature and the austenite start and finish temperature, respectively, which depend on material composition and processing conditions. Austenite, also called parent phase, is stable above $A_f$, whereas multi-variant twinned martensite (product phase) is stable below $M_f$. Under mechanical load, martensite variants align along the load direction assuming a detwinned configuration. The shape memory effect consists of the ability to recover high residual deformations induced below $M_f$, by simply heating the material above $A_f$. Conversely, pseudoelasticity is the ability to recover large stress-induced deformations (up to 7-8\%) above $A_f$ by removing the mechanical load. This work focuses in particular on the pseudoelastic behavior, featuring a characteristic flag-shaped stress-strain response consisting of elastic loading of the austenitic phase, forward phase transformation in detwinned martensite, elastic loading/unloading of the martensitic phase, reverse phase transformation, and elastic unloading with complete strain recovery. 

Due to their remarkable properties, SMAs have been extensively studied over the years, with several advanced engineering applications including biomedical devices, actuators, energy absorption systems, and vibration damping devices \citep{Duerig1999, Benafan2019}. In particular, the SMA industry has been dominated by products for medical applications, with pseudoelastic Nickel-Titanium (Ni-Ti) SMAs adopted as gold standard for designing self-expanding cardiovascular devices such as peripheral stents and heart valve frames \citep{Duerig1999}. 

The fatigue and fracture behavior of SMAs has received great interest for many years. Indeed, most SMA components during their working life undergo significant cyclic loads, either mechanical as stents, or thermal as actuator springs. Repeated phase transformation cycles result in a progressive decrease in the functional properties of the alloy due to the increase in dislocation density (functional fatigue), and in the nucleation of microcracks that eventually lead to abrupt fracture after a sufficient number of cycles (structural fatigue) \citep{Eggeler2004}. 
The structural fatigue of SMAs has been extensively studied since early applications, mainly adopting an experimental perspective. Earlier studies exploited total life approaches which aimed to prevent crack nucleation in biomedical components of micrometric size; the literature is rich in examples, with \citet{Robertson2012} providing a comprehensive review. The last two decades have seen an increasing interest also in fracture mechanics methods accounting for crack propagation in large SMA components able to sustain significant crack growth while maintaining their functionality, such as dampers and actuators \citep{Baxevanis2015}. 
The crack tip region has been experimentally characterized in polycrystalline Ni-Ti samples by means of in situ synchrotron X-ray diffraction, revealing stress-induced martensitic transformation ahead of the crack which affects the process zone, resulting in a complex failure mechanism \citep{Gollerthan2009}. Phase transformation at the crack tip is responsible for slow and stable crack growth under monotonically increasing load, as it leads to dissipated energy which must be supplied by the external loading to advance the crack \citep{Robertson2007}.

In the past years, computational studies on the fracture process in SMAs were mainly devoted to analyze the phase transformation fields near static cracks or investigate the toughening effect of stress-induced phase transformation associated with quasi-static crack growth. The former class of studies adopted continuum models discretized with the standard finite element method to characterize the extension of the phase transformation region at the crack tip and the resulting stress redistribution for different loading conditions \citep{Wang2005, Wang2007, Baxevanis2012}. The latter class of works investigated steady crack advancement, focusing on the effect of reverse phase transformation in the wake of the growing crack and martensite reorientation on fracture toughening. Most of the published studies are either based on linear elastic fracture mechanics tools, such as the virtual crack closure technique \citep{Baxevanis2013, Baxevanis2016}, non-linear elastic fracture mechanics methods based on the J-integral or similar energetic parameters \citep{Baxevanis2014, Yi2001}, cohesive zone models \citep{Freed2007}, or the extended finite element method \citep{Ardakani2015}.

Among the available modeling strategies, the phase-field approach has gained increasing popularity across a wide range of applications in computational mechanics and materials science. In general, the phase-field framework is adopted for modeling systems characterized by sharp discontinuities at interfaces of unknown location by introducing an additional field variable which interpolates between different states through a smooth transition. 
In the context of superelastic SMAs, phase-field formulations and related gradient-based regularization have been successfully employed to describe phase transformation phenomena and their possible localization. Notable examples include the works by \citet{Pham2014} and \citet{Baldelli2015}, where a softening behavior is introduced to capture localized phase transformation at the macroscopic scale, following an energetic approach. Fracture modeling represents one of the most prominent applications of the phase-field approach, where the crack phase-field approximates the sharp crack discontinuity by smoothly transitioning between the intact and the fully damaged material states. 
What makes this approach particularly attractive is its ability to simulate complex processes such as crack initiation, propagation, merging, and branching \citep{Ambati2015}. Growing cracks are automatically tracked by the evolution of the phase-field on a fixed mesh, allowing to deal with complex three-dimensional crack patterns without the need for re-meshing or ad hoc additional criteria \citep{Gerasimov2019}.
The phase-field approach was originally proposed for brittle fracture by \citet{Bourdin2000} as the regularization of the variational formulation by \citet{Francfort1998} recasting Griffith's energetic fracture criterion as the minimization problem of an energy functional. 
Later, the phase-field framework was re-interpreted as a special type of gradient damage models characterized by an internal length, which converge to Griffith's model (in the sense of the so-called $\Gamma$-convergence) for the internal length tending to zero \citep{Pham2011b}. Accordingly, the terms ``phase-field variable'' and ``damage variable'' are generally adopted equivalently in the literature. The characteristic features of gradient damage models are \citep{Alessi2014, Alessi2015}: i) a material stiffness decreasing with damage to zero for the fully damaged state, ii) the presence of a critical damage stress that decreases to zero with increasing damage, and iii) a gradient damage term in the energy formulation that limits damage localization through an internal length parameter. Consequently, in a one-dimensional (1D) setting, the crack nucleation process develops as follows \citep{Alessi2014, Alessi2015}: damage first occurs in a point of the bar where the critical stress is reached; then, due to the softening character of the model, damage localizes in a zone whose width is related to the internal length; finally, when the maximum damage reaches its ultimate value, a crack appears at the center of the damage zone. The energy dissipated during this process is related to the critical energy release rate of Griffith's theory $G_c$.
After having been widely applied to brittle fracture \citep{Ambati2014, Kuhn2010, Miehe2010, Gerasimov2018}, the phase-field method has been extended to numerous other cases, including ductile fracture \citep{Alessi2014, Alessi2015, Alessi2018, Ambati2015}, composite delamination \citep{Quintanas2019}, hydrogen embrittlement \citep{Paneda2018}, and fatigue \citep{Carrara2020} to name a few. 

Recently, \citet{Simoes2021}, \citet{Simoes2022} proposed a phase-field formulation for fatigue and fracture in SMAs adopting the constitutive model by \citet{Lagoudas2008} and the one developed by \citet{Auricchio1997b}, combined with a phase-field theory \citep{Bourdin2000}. Several boundary value problems were addressed, analyzing the stress-induced phase transformation at the crack tip and the resulting energy dissipation and material toughening. Furthermore, the fatigue degradation function proposed by \citet{Carrara2020} was incorporated in the energy functional to address fatigue effects in different geometries: a single edge notched specimen, a Ni-Ti stent, and a three-dimensional diamond lattice structure. However, the predictive capabilities of the model were not assessed with experimental data.

\citet{Hasan2022} aimed to improve the formulation by \citet{Simoes2021}, \citet{Simoes2022} to obtain a flexible finite-strain framework able also to address actuation loading and martensite reorientation. In this case, the fracture process was assumed to be driven by the elastic strain energy only, instead of the total energy as in \citet{Simoes2021}, \citet{Simoes2022}, supposing that the inelastic deformation contributes to fracture only through stress redistribution. The analysis was restricted to monotonic loading conditions. The authors were able to reproduce experimental observations under isothermal and isobaric loading, both in simple two-dimensional samples and complex three-dimensional structures, such as the toughening effect leading to stable crack growth.

In this work, we develop a new phase-field model for fracture and fatigue in SMAs in a 1D setting which, unlike the previous ones, preserves a variational nature with its related advantages \citep{vicentini2024energy}. The starting point of our study is the work in \citet{Alessi2014, Alessi2015, Alessi2018}, on gradient damage coupled with plasticity. Starting from a brittle gradient damage model, the authors introduce plastic strains, coupling their evolution with damage. This interestingly leads to the nucleation of cohesive cracks, namely, cracks characterized by a discontinuity in displacement for non-vanishing stress. In a similar way, in our work damage is coupled with the so-called transformation strain resulting from the SMA phase transformation. 
The coupling is introduced by assuming the dependence of the material parameters governing the onset of phase transformation on damage. Once the total energy is defined, the evolution problem is solved by assuming damage irreversibility and applying the stability and energy balance principles. In particular, the damage and transformation yield criteria are obtained from first-order stability, while the consistency equations and the transformation flow rules are recovered from the energy balance. 
From a numerical point of view, the variational approach is suitable for numerically solving the evolution problem through an alternate minimization algorithm. Accordingly, the total energy of the discretized domain is minimized with respect to one single discrete state field at each step, keeping the other fixed, and iterating until convergence.

Moving a step forward with respect to \citet{Alessi2014, Alessi2015, Alessi2018}, the present work additionally aims to address fatigue damage effects in order to establish a predictive framework suitable for real SMA applications.
In particular, we focus on cardiovascular applications of SMAs, where the material works in the pseudoelastic regime at fixed temperature and is subjected to mechanical cyclic loads related to blood pulsation or body movements. 
In this context, due to the limited size of the structural features (100--200 \textmu m), fatigue performance is primarily governed by crack nucleation, as once a macroscopic crack nucleates, it propagates rapidly and leads to failure shortly afterwards \citep{Robertson2012}. Accordingly, we consider a simplified 1D setting, modeling the crack nucleation process, and we investigate whether the proposed phase-field model can provide quantitative estimates of the fatigue life of wires or other uniaxial samples commonly used to characterize the fatigue behavior of Ni-Ti SMAs in cardiovascular contexts, subjected to tension-tension loading conditions. This scenario is well supported by experimental data in the literature, which can be leveraged to critically assess the predictive capability of the model. A comparison between numerical predictions and experimental results in such a simplified setting is regarded as a key step to evaluate the potential of the proposed approach and to assess the merit of extending it to more complex three-dimensional scenarios. Therefore, validation against experimental fatigue data represents a particular strength of this work, as such direct comparisons are not performed in existing studies.

The paper is organized as follows. In Section \ref{sec2}, we summarize the SMA constitutive model adopted in this study, recasting it in a variational form. In Section \ref{sec3}, we present a phase-field model coupling phase transformation and damage, following energetic principles to derive the governing equations of the evolution problem. Section \ref{sec4} analyses the monotonic response. First, we describe analytically the homogeneous response, i.e. the response of a material point subjected to a uniaxial monotonic strain. Then, we address the localization response of a bar subjected to a monotonically increasing elongation. This problem is solved numerically and the implemented algorithm is described in detail. Fatigue effects are studied in Section \ref{sec5}, describing how the model naturally tackles the evolution of damage under repeated loads. We address both the homogeneous and the localization responses, providing several numerical examples. Finally, a real case study is presented in Section \ref{sec6}, considering Ni-Ti multi-wire samples used to characterize material fatigue in the SMA cardiovascular industry. The model is calibrated on the basis of uniaxial static data. Then, uniaxial fatigue tests in several loading conditions are simulated, evaluating the ability of the model in predicting the fatigue life in comparison with experimental results.
\section{Variational formulation of a 1D constitutive model for SMAs} 
\label{sec2}
Among the constitutive models proposed in the literature to describe the SMA response, we consider the thermomechanical model originally proposed by \citet{Souza1998} and revisited by \citet{Auricchio2004}.
In a 1D setting, we assume a decomposition of the scalar strain $\varepsilon$ into an elastic component $\varepsilon^e$ and an inelastic component, the transformation strain $e^{tr}$. The model requires $0\leq \lvert e^{tr}\rvert \leq \varepsilon_L $, with $\varepsilon_L$ as the maximum transformation strain measured at the end of phase transformation during a uniaxial test. When $e^{tr}=0$ the material is in its parent phase, while for $\lvert e^{tr} \rvert=\varepsilon_L$  the material is fully transformed into the product phase. For intermediate conditions, the material consists of a mixture of parent and product phases. The following simplifying assumptions are introduced: (i) small strains, (ii) the temperature is fixed at a constant value (as we focus on pseudoelastic behavior only), and (iii) a symmetric behavior between tension and compression (since we only model tensile tests). 

Assuming the total strain $\varepsilon$ as control variable and the transformation strain $e^{tr}$ as internal variable, the total energy density is expressed as
\begin{equation}\label{eq1}
\begin{split}
& W_{T}(\varepsilon, e^{tr}, \overline{e^{tr}}) = \\ & W_{el}(\varepsilon - e^{tr}) + W_{ch}(e^{tr}) + W_{tr}(e^{tr}) + W_{d}(\overline{e^{tr}}).
\end{split}
\end{equation}

The energy contributions appearing in (\ref{eq1}) are defined as follows:
\begin{itemize}
    \item[-] the elastic strain energy density $W_{el}$ is
    \begin{equation} \label{eq2}
    W_{el} = \frac{1}{2}  E_0 (\varepsilon - e^{tr})^{2}
    \end{equation}
    with $E_0$ as the elastic modulus accounting for the mixture of material phases;
    \item[-] the chemical energy density $W_{ch}$ due to the thermally-induced martensitic transformation is
    \begin{equation} \label{eq3}
    W_{ch} = \tau_{M_0}\lvert e^{tr}\rvert
    \end{equation}
    with $\tau_{M_0}$ as a material constant equal to $\beta \langle T - M_{f}\rangle$, where $\beta$ is the coefficient of the linear relation between the transformation stress and the temperature, and $\langle \bullet \rangle$ denotes the positive part of the argument. In a 1D setting, $\tau_{M_0}$ can also be interpreted as the center of the elastic domain for $\lvert e^{tr}\rvert \to 0$;
    \item[-] the transformation strain energy density $W_{tr}$ due to the transformation-induced hardening is 
    \begin{equation} \label{eq4}
    W_{tr} = \frac{1}{2} h_0 \lvert e^{tr}\rvert^{2}
    \end{equation}
    where $h_0$ is a material parameter representing the slope of the linear relation between stress and transformation strain in the uniaxial case;
    \item[-] the dissipation energy density $W_{d}$ is  
    \begin{equation} \label{eq5}
    W_{d} = R_0 \overline{e^{tr}} 
   \end{equation}
    with $R_0$ as a material parameter defining the radius of the elastic domain and $\overline{e^{tr}}$ representing the accumulated transformation strain 
    \begin{equation} \label{eq6}
    \overline{e^{tr}}=\int_{0}^{t}   \lvert \dot{e}^{tr}(\tau)\rvert d\tau.
     \end{equation}
\end{itemize}
The constitutive equations of the model are formulated applying the following three conditions:
\begin{equation} \label{eq7}
\text{Stress-strain relation:     } \sigma = \frac{\partial W_T}{\partial \varepsilon},  
\end{equation}
\begin{equation} \label{eq8}
\begin{gathered}
\text{Stability condition:} \\
W_T(\varepsilon, e^{tr}, \overline{e^{tr}})  \leq W_T(\varepsilon, e^{tr^*}, \overline{e^{tr}} + \lvert e^{tr^*} - e^{tr}\rvert ), \\ \forall e^{tr^*} \in \mathbb{R}, 
\end{gathered}
\end{equation}
\begin{equation} \label{eq9}
\text{Energy balance:     } \dot W_T = \sigma \dot \varepsilon,
\end{equation}
where a superposed dot denotes differentiation with respect to time. While (\ref{eq7}) is equivalent to the stress-strain relationship
\begin{equation}  \label{eq10}
\sigma (\varepsilon, e^{tr}) = E_0 (\varepsilon - e^{tr}),
\end{equation}
it can be proven, as shown in Appendix \ref{appendix:A}, that (\ref{eq8}) and (\ref{eq9}) are equivalent to the transformation yield criterion
\begin{equation}  \label{eq11}
f = \lvert X \rvert -R_0 \leq 0,
\end{equation} 
with
\begin{equation}  \label{eq12}
X = \sigma - (\tau_{M_0} + h_0 \lvert e^{tr} \rvert) \frac{\partial \lvert e^{tr}  \rvert}{\partial e^{tr}}
\end{equation}
representing the so-called transformation stress, and to the evolution equation for the transformation strain 
\begin{equation} \label{eq13}
\dot e^{tr} = \dot \lambda \,\text{sign} (X)
\end{equation}
with the Kuhn-Tucker conditions
\begin{equation}  \label{eq14}
f \leq 0, \ \ \ \ \ \ \dot \lambda \geq 0, \ \ \ \ \ \  \dot \lambda f = 0,
\end{equation}
with $\lambda$ playing a role similar to that of a plastic multiplier in a plasticity model. For $f<0$, the material behaves elastically and the transformation strain rate is equal to zero. When the material undergoes phase transformation, that is for $f=0$, the transformation strain increases during the forward phase transformation ($X>0)$ and decreases during the reverse phase transformation ($X<0$). 
\section{Variational phase-field model of phase transformation coupled with damage in SMA wires}
\label{sec3}
We now develop a 1D variational phase-field model that combines damage evolution with phase transformation occurring in SMAs, by coupling with damage the constitutive model of SMAs discussed in the previous section following the rationale of previous works on damage-plasticity coupling \citep{Alessi2014, Alessi2015,Alessi2018}.
The model is developed in a rate-independent and quasi-static framework.
\subsection{State variables and energetic quantities}
Consider the 1D straight bar whose reference configuration is the interval $[0, L]$. The end $x=0$ is fixed, while the end $x=L$ is subjected to a time-dependent displacement $U(t)$. 
The scalar state variables at each point of the body $x \in [0,L]$ are the displacement field $u(x,t)$, related to the infinitesimal total strain $\varepsilon (x,t)$, the transformation strain $e^{tr} (x,t)$, the accumulated transformation strain $\overline{e^{tr}}(x,t)$ defined by (\ref{eq6}), the damage variable $\alpha(x,t)$, and the damage spatial gradient $\alpha'(x,t):= \frac{\partial \alpha(x, t)}{\partial x}$. The scalar damage field $\alpha$ is assumed to be bounded in $[0,1]$, with $\alpha=0$ corresponding to the undamaged state and $\alpha=1$ to the fully damaged state, and must satisfy the irreversibility condition $\dot \alpha \geq 0$ on $[0,L]$.
It is additionally assumed that at time $t=0$ the material is undeformed, undamaged, and not transformed (i.e. in the austenitic state).
The total energy density is assumed to be
\begin{equation} \label{eq15}
\begin{split}
&W(\varepsilon, e^{tr}, \overline{e^{tr}},\alpha, \alpha') = \frac{1}{2}  E(\alpha) (\varepsilon - e^{tr})^{2} \\&+ \tau_{M}(\alpha)\lvert e^{tr}\rvert + \frac{1}{2} h(\alpha) \lvert e^{tr}\rvert^{2} + R(\alpha) \overline{e^{tr}} \\& + w(\alpha) + w_{1}l^{2}(\alpha')^{2}.
\end{split}
\end{equation}
In (\ref{eq15}), the first four contributions come from the SMA constitutive model, whereas the last two terms are associated to a well-known phase-field (or gradient damage) model for brittle fracture \citep{Pham2011b}. 
The state function $E(\alpha)$, representing the elastic modulus of the material, is assumed to be sufficiently smooth and monotonically decreasing with $\alpha$ up to zero. 
The material parameters $\tau_{M}$, $h$, and $R$ of the SMA constitutive model are now expressed as functions of the damage state. Assuming a softening behavior, $\tau_M(\alpha)$, $h(\alpha)$, and $R(\alpha)$ are required to be monotonically decreasing with damage up to zero for the fully damaged state. Therefore, we assume:
\begin{equation} \label{eq16_0}
\begin{split}
& E(0)=E_{0} >0, \hspace{1.9cm} E(1)=0, \\&  E'(\alpha)<0 \ \text{for} \ \alpha=[0,1), \hspace{0.5cm} E'(1) \leq 0;
\end{split}
\end{equation}
\vspace{-\baselineskip}
\begin{equation}  \label{eq16}
\begin{split}
&\tau_{M}(0)=\tau_{M_0} >0, \hspace{1.5cm} \tau_{M}(1)=0,  \\& \tau_{M}'(\alpha)<0 \ \text{for} \ \alpha=[0,1), \hspace{0.4cm} \tau_{M}'(1) \leq 0;
\end{split}
\end{equation}
\begin{equation}  \label{eq17}
\begin{split}
& h(0)=h_0 >0, \hspace{2cm}  h(1)=0,  \\&  h'(\alpha)<0  \ \text{for} \ \alpha=[0,1), \ \hspace{0.5cm} h'(1) \leq 0;
\end{split}
\end{equation}
\begin{equation}  \label{eq18}
\begin{split}
& R(0)=R_0 >0, \hspace{1.9cm} R(1)=0,  \\& R'(\alpha)<0 \ \text{for} \ \alpha=[0,1), \hspace{0.5cm}  R'(1) \leq 0. 
\end{split}
\end{equation}
The scalar damage function $w(\alpha)$ stands for the density of the energy dissipated by the material during a homogeneous damage process, with $w_1=w(1)$ denoting the specific fracture energy. We assume that it satisfies the following conditions:
\begin{equation}  \label{eq19}
w(0)=0, \ \ \ \ \ \ w'(\alpha)>0.
\end{equation}
Finally, the last term in (\ref{eq15}) is the standard gradient regularization, with $l>0$ representing the internal length, or regularization length, related to the width of the localized damage profile.

The total energy of the bar (at a fixed time $t$ which we omit to indicate explicitly) can be expressed as the following functional of the global state field $\xi=(u, e^{tr},  \overline{e^{tr}}, \alpha)$
\begin{equation}  \label{eq24}
\begin{split}
&\mathcal{W} (u, e^{tr},  \overline{e^{tr}}, \alpha) = \\&\int_{0}^{L} W(\varepsilon(u(x)), e^{tr}(x),  \overline{e^{tr}}(x), \alpha(x), \alpha'(x)) dx.
\end{split}
\end{equation}
The above expression of the energy makes sense provided that the global state field is smooth enough. For the damage variable, in a 1D setting the gradient term requires that $\alpha \in C^0([0,L])$, meaning that damage must be continuous, but its derivative can be discontinuous. The displacement and the transformation strain fields may not always be smooth due to the localization of deformation induced by the softening behavior of the model \citep{Alessi2014}. 
Following \citet{Alessi2014, Alessi2015, Alessi2018}, we thus assume that the global state field $\xi$ is only piecewise smooth and that its singular part is localized on a jump set $J(u) \in [0,L]$, which contains a finite number of points. By hypothesis, the jump set has no intersection with the boundaries $x=0$ and $x=L$. 
We assume that the displacement field $u$ is continuously differentiable on the regular domain $[0,L] \backslash J(u)$ and admits a jump discontinuity on $J(u)$. Accordingly, the total strain, the transformation strain and the accumulated transformation strain can be decomposed into their regular and singular parts, respectively indicated as $(\bullet)^R$ and $(\bullet)^S$. The total strain can be written as
\begin{equation}  \label{eq25}
\varepsilon=\varepsilon^{R} + \varepsilon^{S} = u'(x) + \sum_{x_i \in J(u)} [\![ u ]\!](x_i) \delta_{x_i},
\end{equation}
where 
$\delta$ is the Dirac measure concentrated on the singular points $x_i \in J(u)$. In a similar way, the transformation strain and the accumulated transformation strain fields are decomposed as
\begin{equation}  \label{eq26}
e^{tr}=e^{tr^R} + e^{tr^S} = e^{tr^R}(x) + \sum_{x_i \in J(u)} [\![ u ]\!](x_i) \delta_{x_i},
\end{equation}
\begin{equation}  \label{eq27}
\overline{e^{tr}}=\overline{e^{tr^R}} + \overline{e^{tr^S}} = \overline{e^{tr^R}}(x) + \sum_{x_i \in J(u)} \overline {[\![ u ]\!]}(x_i) \delta_{x_i},
\end{equation}
with 
\begin{equation}  \label{eq28}
 \overline {[\![ u ]\!]} (x_i)=\int_{0}^{t}\lvert [\![ \dot u ]\!](x_i,\tau) \rvert d\tau, \ \ \ \forall x_i \in J(u). 
\end{equation}
Note that the regular part of the transformation strain is required to be at least continuous on $[0,L] \backslash J(u)$, and that its singular part is the same as the singular part of the total strain field in order for the elastic energy to be finite. Moreover, the number of singular points can only increase in time, meaning that if a displacement jump develops at a certain point $x_i$ at time $t_i$, then $\overline {[\![ u ]\!]}(x_i,t) > 0$ for all $t \geq t_i$. Finally, with these assumptions, the total energy of the bar at a fixed time $t$ for the global field $\xi=(u, e^{tr},  \overline{e^{tr}}, \alpha)$ can be rewritten as
\begin{equation}  \label{eq29}
\begin{split}
&\mathcal{W} (u, e^{tr},  \overline{e^{tr}}, \alpha) =\\& \int_{[0,L]\backslash J(u)} \biggl[ \frac {1}{2}E(\alpha(x)) (u'(x) - e^{tr^R}(x))^{2} \\& + \tau_{M}(\alpha(x))\lvert e^{tr^R}(x)\rvert + \frac{1}{2} h(\alpha(x)) \lvert e^{tr^R}(x)\rvert^{2} \\ 
&+ R(\alpha(x)) \overline{e^{tr^R}}(x) + w(\alpha(x)) + w_{1}l^{2}(\alpha'(x))^{2}\biggr] dx\\
& + \sum_{x_i \in J(u)} \biggl[ \tau_{M}(\alpha(x_i)) \lvert [\![ u ]\!](x_i) \rvert + \frac{1}{2} h(\alpha(x_i)) \lvert [\![ u ]\!](x_i) \rvert^2 \\&+ R(\alpha(x_i)) \overline {[\![ u ]\!]}(x_i)\biggr].
\end{split}
\end{equation}
\subsection{The 1D evolution problem}
The 1D evolution problem is formulated adopting a variational approach, constructing the solution as the process $t\mapsto \xi_t = (u_t, e^{tr}_t, \overline{e^{tr}}_t, \alpha_t)$ (where we denote time dependency with the subscript $t$) which satisfies the irreversibility condition, the stability condition, and the energy balance.
\begin{itemize}
\item \textit{Irreversibility condition} \\
It requires that at every point the damage can only increase in time
\begin{equation} \label{eq30}
\dot \alpha_t(x) \geq 0, \ \ \ \ 0 \leq \alpha_t(x) < 1, \ \ \ \ \forall x \in [0,L].
\end{equation}
\item \textit{Stability condition} \\
Let $\xi_t = (u_t, e^{tr}_t, \overline{e^{tr}}_t, \alpha_t)$ be the state of the bar at time $t$  and consider $\xi^* =(u^*, e^{tr^*}, \overline{e^{tr}}^*, \alpha^*)$ as the virtual state $\xi^* = \xi_t + a(v, q, \lvert q \rvert, \beta)$, where $a>0$ and $v$, $q$, $\beta$ are perturbations respectively on displacement, transformation strain, and damage. For $u^*$ to be kinematically admissible, the field $v$ must be such that $v(0)=v(L)=0$. Moreover, $v$ is assumed piecewise smooth, with discontinuities localized on the set of points $J(v)$. For the perturbed elastic strain to be integrable, $q$ must be such that $q = q^R + [\![ v ]\!] \delta_{J_v} $. Lastly, in order that $\alpha_t \leq \alpha^* < 1$ and $\alpha^* \in H^1([0,L])$, it is necessary and sufficient that $\beta \geq 0$, $\beta \in H^1([0,L])$ and $a$ be small enough. The state $\xi_t = (u_t, e^{tr}_t, \overline{e^{tr}}_t, \alpha_t)$ is locally stable if, for any admissible perturbation $(v, q, \beta)$, there exists $\overline a >0$ such that for all $a \in [0, \overline a]$
\begin{equation} \label{eq31}
\begin{split}
&\mathcal{W} (u_t + av, e^{tr}_t + aq,  \overline{e^{tr}}_t + a\lvert q \rvert , \alpha_t + a\beta) \\& \geq \mathcal{W} (u_t, e^{tr}_t,  \overline{e^{tr}}_t, \alpha_t).
\end{split}
\end{equation}
\item \textit{Energy balance} \\
It requires that the variation of the total energy of the bar is equal to the power of the external force acting at the end of the bar $x=L$
\begin{equation} \label{eq32}
\frac{d}{dt} \mathcal{W} (u_t, e^{tr}_t,  \overline{e^{tr}}_t, \alpha_t) = \sigma_t(L) \dot U_t.
\end{equation}
\end{itemize}
The three above energetic principles provide as necessary conditions the evolution laws summarized in Table \ref{tab1}. Their detailed derivation is reported in Appendix \ref{appendix:B}. Note that the damage consistency equation and the transformation flow rule hold both on the regular and the singular parts of the domain. The damage yield function on the regular domain contains a gradient term while on the singular domain it introduces a relation between the displacement jump and the jump of the damage gradient through the singular part of the transformation strain. Therefore, a discontinuity of the transformation strain generally induces a discontinuity of the damage gradient and vice versa.
\begin{table*} [t]
\fontsize{8pt}{8pt}
\centering
\caption{Governing equations of the evolution problem obtained from irreversibility, stability, and energy balance. We omit the spatial and temporal dependence of the state variables to simplify the notation and we introduce the compliance state function $S(\alpha) = E^{-1}(\alpha)$.}
\begin{tabular}{c c} 
\hline
\multicolumn{2}{c}{Regular domain $[0,L]\backslash J(u)$} \\ 
\hline
\\ Equilibrium equation  & $\sigma'=0$ \ \ \ with $\sigma(\varepsilon, e^{tr}, \alpha)=E(\alpha) (\varepsilon - e^{tr})$\\ \\
Transformation yield criterion & $\lvert X(\sigma,e^{tr},\alpha) \rvert - R(\alpha)  \leq 0$ \ \ \ with $X(\sigma,e^{tr},\alpha) = \sigma - (\tau_{M}(\alpha) + h(\alpha) \lvert e^{tr} \rvert) \frac{\partial \lvert e^{tr}  \rvert}{\partial e^{tr}}$\\ \\
Transformation flow rule & 
\begin{math}
\dot {e}^{tr} \left\{ \begin{array}{rcl} \geq 0 & \mbox{if} & X(\sigma,e^{tr},\alpha) = R(\alpha) \\ =0 & \mbox{if} & \lvert X(\sigma,e^{tr},\alpha) \rvert -R(\alpha) < 0 \\ \leq 0 & \mbox{if} & X(\sigma,e^{tr},\alpha) = -R(\alpha) \end{array}\right.
\end{math} \\ \\
Damage yield criterion &  
\begin{math}
\frac{1}{2}S'(\alpha)\sigma^2 - \tau_M'(\alpha) \lvert e^{tr} \rvert - \frac{1}{2} h'(\alpha) \lvert e^{tr} \rvert^2 - R'(\alpha) \overline {e^{tr}} - w'(\alpha) + 2w_1 l^2 \alpha'' \leq 0
\end{math} \\ \\
Damage consistency equation &  
\begin{math}
\biggl( \frac{1}{2}S'(\alpha)\sigma^2  - \tau_M'(\alpha) \lvert e^{tr} \rvert - \frac{1}{2} h'(\alpha) \lvert e^{tr} \rvert^2 - R'(\alpha) \overline {e^{tr}} - w'(\alpha) + 2w_1 l^2 \alpha'' \biggr) \dot \alpha =0 
\end{math} \\ \\
\hline
\multicolumn{2}{c}{Singular domain $J(u)$} \\
\hline
\\ Equilibrium equation & $[\![ \sigma ]\!]=0 $  \\ \\
Transformation yield criterion & $\lvert X(\sigma,[\![u]\!],\alpha) \rvert - R(\alpha) \leq 0$  \\ \\ 
Transformation flow rule & 
\begin{math}
  {[\![\dot u]\!]} \left\{ \begin{array}{rcl} \geq 0 & \mbox{if} & X(\sigma,[\![u]\!],\alpha) = R(\alpha) \\ =0 & \mbox{if} & \lvert X(\sigma,[\![u]\!],\alpha) \rvert -R(\alpha) < 0 \\ \leq 0 & \mbox{if} & X(\sigma,[\![u]\!],\alpha) = -R(\alpha) \end{array}\right.
\end{math} \\ \\ 
Damage yield criterion & \begin{math}
2w_1 l^2 [\![ \alpha' ]\!] -\tau_M'(\alpha) \lvert [\![ u ]\!] \rvert - \frac{1}{2} h'(\alpha) \lvert [\![ u ]\!] \rvert^2 - R'(\alpha) \overline {[\![ u ]\!]}  \leq 0
\end{math} \\ \\
Damage consistency equation &   
\begin{math}
\biggl( 2w_1 l^2 [\![ \alpha' ]\!] - \tau_M'(\alpha) \lvert [\![ u ]\!] \rvert - \frac{1}{2} h'(\alpha) \lvert [\![ u ]\!] \rvert^2 - R'(\alpha) \overline {[\![ u ]\!]}  \biggr)  \dot \alpha = 0
\end{math} \\ \\
\hline
\multicolumn{2}{c}{Boundary conditions $\{0,L\}$ }\\
\hline
\\ Dirichlet boundary conditions &  $u(0)=0, \ \ u(L)=U(t)$ \\ \\
\multirow{2}{13em}{Damage boundary conditions} &   $\alpha'(0) \leq0, \ \ \alpha'(L) \geq0$ \\
& $\alpha'(0) \dot \alpha(0) =0, \ \ \alpha'(L) \dot \alpha(L) =0$ \\ \\
\hline
\end{tabular}
\label{tab1}
\end{table*}
\subsection{Constitutive equations}
Among the choices proposed in the literature, we consider the one-parameter class of models presented in \cite{Alessi2018}. The adopted constitutive functions are defined as
\begin{equation}  \label{eq33}
\begin{split}
&E(\alpha) = (1-\alpha)^2 E_0, \  \  w(\alpha)=w_1 \alpha \\
&\tau_M(\alpha)=(1-\alpha)^s \tau_{M_0}, \ \ h(\alpha)=(1-\alpha)^s h_0, \\&  R(\alpha)=(1-\alpha)^s R_0,
\end{split}
\end{equation}
where $E_0$, $w_1$, $\tau_{M_0}$, $h_0$, and $R_0$ are parameters of the undamaged material. The constitutive equations $\tau_M(\alpha)$, $h(\alpha)$, and $R(\alpha)$  have the same form as the yield stress function in the ductile fracture model by \cite{Alessi2018}. The parameter $s>0$ governs the softening behavior of the stress during damage, and the conditions (\ref{eq16_0})-(\ref{eq19}) are automatically satisfied. In the following, it will be shown that the specific choice of the parameters $(\tau_{M_0}, h_0, R_0, w_1, s)$ originates different material responses. Specifically, given the formulation of the damage constitutive function $w(\alpha)$, the response will always be characterized by an initial elastic stage (E), which may be followed by a stage of damage evolution only (D), transformation strain evolution only (T), or coupled transformation strain and damage evolution (TD).
\section{Monotonic response}
\label{sec4}
In this section, we study the response of the model presented in Section \ref{sec3} under monotonic loading, first at the level of a material point (homogeneous response) and then at the structural level for a 1D bar subjected to a monotonically increasing imposed end displacement. 
\subsection{Homogeneous response}
\label{sec4-1}
In the following, we discuss the homogeneous material response, which implies considering either a spatial domain with uniform damage and strain fields or simply a material point. We consider a monotonically increasing total strain parametrized in terms of the time variable $t$, $\varepsilon=t$. The material is assumed initially undeformed, not transformed, and undamaged:
\begin{equation} \label{eq34}
\xi_0 = (\varepsilon_0, e^{tr}_0, \overline{e^{tr}}_0, \alpha_0) = (0,0,0,0).
\end{equation}
Since under monotonic loading at each time instant $e^{tr}_t = \overline{e^{tr}}_t$, the only unknowns are the transformation strain and the damage variable.
The material response can be derived exploiting the stress-strain relation $\sigma=E(\alpha) (\varepsilon - e^{tr})$ and the evolution laws summarized in Table \ref{tab1}, neglecting the spatial derivatives and considering the regular domain only. The transformation yield criterion provides the stress at the onset of phase transformation
\begin{equation} \label{eq35}
\sigma_T (e^{tr}, \alpha) = \left\{ 
\begin{aligned} &\sigma^{fwd}_T (e^{tr}, \alpha) =\\ &(1-\alpha)^s \biggl((\tau_{M_0} + h_0 \lvert e^{tr} \rvert) \frac{\partial \lvert e^{tr}  \rvert}{\partial e^{tr}} +R_0 \biggr) \\& \mbox{if} \ \ X(\sigma,e^{tr},\alpha) = R(\alpha) \\ \\
&\sigma^{rev}_T (e^{tr}, \alpha) =\\& (1-\alpha)^s  \biggl((\tau_{M_0} + h_0 \lvert e^{tr} \rvert) \frac{\partial \lvert e^{tr}  \rvert}{\partial e^{tr}} -R_0  \biggr) \\& \mbox{if} \ \  X(\sigma,e^{tr},\alpha)= -R(\alpha) 
\end{aligned}\right.,
\end{equation}
with $\sigma^{fwd}_T (e^{tr}, \alpha)$ and $\sigma^{rev}_T (e^{tr}, \alpha)$ defining the stress at the onset of the forward and of the reverse transformation, respectively. The damage yield criterion in Table \ref{tab1} provides the damage yield stress as
\begin{equation} \label{eq36}
\begin{split}
&\sigma_D (e^{tr},\overline{e^{tr}}, \alpha)=\\
&\sqrt{
  \begin{aligned}
  E_0 &\biggl{[}  w_1 - s(1-\alpha)^{s-1} \biggl( \tau_{M_0} \lvert e^{tr} \rvert + \frac{1}{2} h_0 \lvert e^{tr} \rvert^2 \\&+  R_0 \overline{e^{tr}} \biggr)  \biggr] (1-\alpha)^3
  \end{aligned}
}.
\end{split}
\end{equation}

We introduce the following simplifying assumptions: (i) $E_0=1$ for any mixture of parent and product phases, and (ii) $\varepsilon_{L}\rightarrow\infty$, such that a fully transformed material state is never attained. Lastly, since in real applications of SMAs the phase transformation stage is expected to always precede the onset of damage, we only consider combinations of material parameters leading to $\sigma_D(0,0,0) > \sigma^{fwd}_T(0,0)$, as reported in Table \ref{tab2}. In the E-T-D model, after the initial elastic stage, a transformation stage and a damage stage occur in sequence; in the E-T-TD model, after the initial elastic stage, the transformation stage is followed by a stage where the evolutions of transformation strain and damage are coupled.
\begin{table} [t!]
\centering
\caption{Combinations of material parameters for studying the monotonic response, with indication of the associated behavior.}
\begin{tabular}{c c c c c c}
\hline
Model  & $\tau_{M_0}$ &  $h_0$  &  $R_0$ &  $w_1$ &  $s$\\
\hline
E-T-D & 0.8 &  0.1&  0.2& 2 &  1 \\
E-T-TD & 0.8 &  0.1&  0.2& 3 &  2 \\
\hline
\end{tabular}
\label{tab2}
\end{table}
\subsubsection{E-T-D model}
\label{sec4-1-1}
\begin{figure*}[t!]
    \centering
    \includegraphics[width=\textwidth]{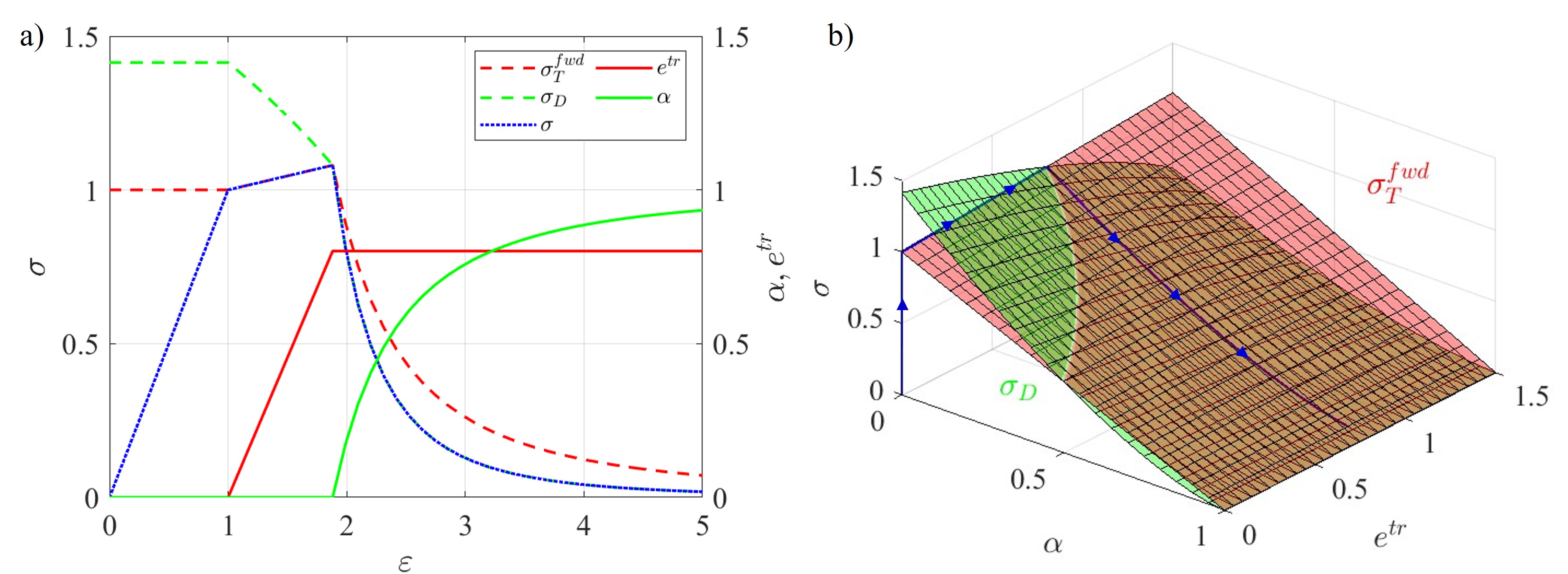}
    \caption{Homogeneous response for the E-T-D model under monotonically increasing strain: a) evolution of the forward transformation stress, damage stress, stress, transformation strain, and damage; b) forward transformation (red) and damage (green) yield surfaces with the evolution path depicted in blue.}
    \label{fig2}
\end{figure*}

Since $\sigma_D(0,0,0)=\sqrt{2} > \sigma^{fwd}_T(0,0)=1$, after the initial elastic stage, a phase transformation stage occurs at the yield point $(\varepsilon,\sigma)=(1,1)$. During the T stage, the transformation strain evolves according to 
\begin{equation} \label{eq37}
    e^{tr}=\frac{E_0 \varepsilon - R_0 - \tau_{M_0}}{E_0 + h_0},
\end{equation}
obtained from the equivalence between the stress-strain relation $\sigma=E_0(\varepsilon - e^{tr})$ and $\sigma^{fwd}_T (e^{tr}, 0)$. Since the damage stress decreases with the transformation strain, at some time damage is triggered, and this time is determined by imposing $\sigma^{fwd}_T (e^{tr}, 0)=\sigma_D (e^{tr}, e^{tr}, 0)$. Then, since the transformation yield stress decreases more slowly with damage than the damage stress, only damage evolves, while the transformation strain remains constant. During stage D, due to the choice of the parameter $s=1$, the damage variable is given by  
\begin{equation} \label{eq38}
    \alpha=1-\frac{w_1 - \tau_{M_0} e^{tr}  - \frac{1}{2}h_0 e^{tr^2} - R_0 \overline{e^{tr}}} {E_0(\varepsilon - e^{tr})^2}.
\end{equation}
Note that the symbol of absolute value is omitted since we consider positive quantities. The material response is depicted in Figure \ref{fig2}a which reports the evolution of the stress, forward transformation stress, damage stress, transformation strain, and damage variable under a monotonically increasing strain. Moreover, Figure \ref{fig2}b shows the transformation and damage yield surfaces with the evolution path, from which the three response stages can be recognized. 
\subsubsection{E-T-TD model}
\label{sec4-1-2}
\begin{figure*}[t!]
    \centering
    \includegraphics[width=\textwidth]{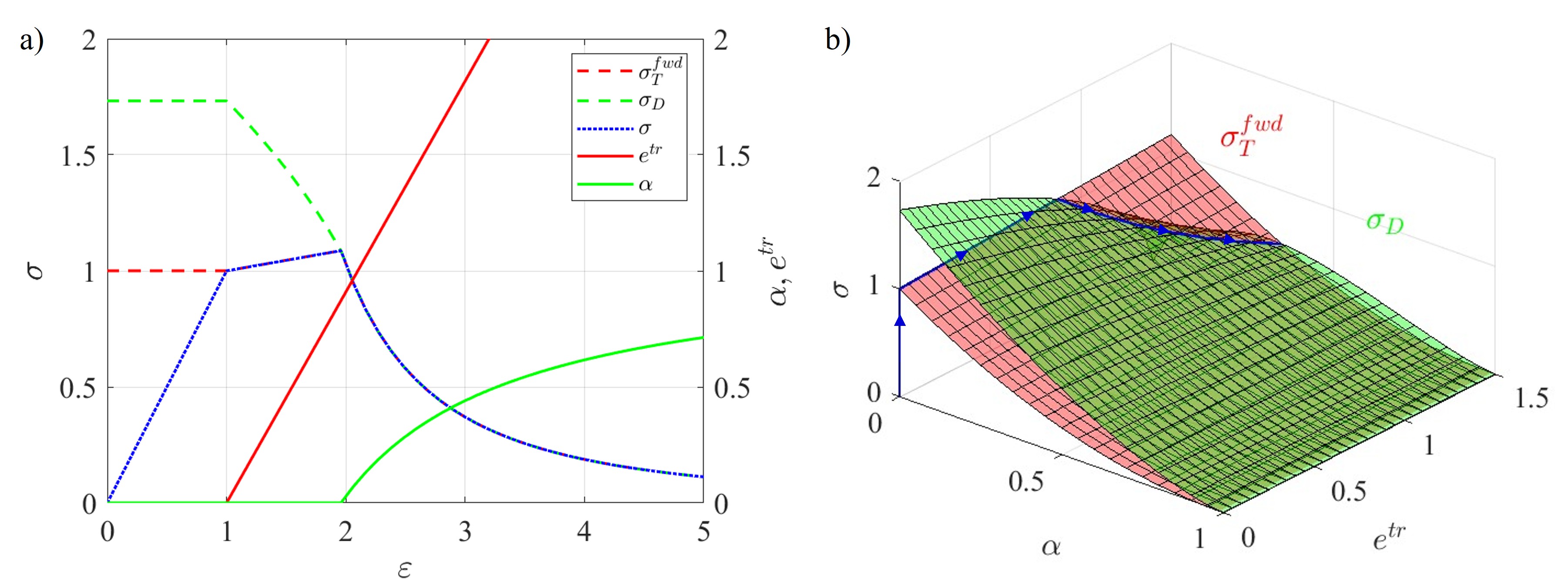}
    \caption{Homogeneous response for the E-T-TD model under monotonically increasing strain: a) evolution of the forward transformation stress, damage stress, stress, transformation strain, and damage; b) forward transformation (red) and damage (green) yield surfaces with the evolution path depicted in blue.}
    \label{fig3}
\end{figure*}

As for the E-T-D model, after the initial elastic stage, a phase transformation stage occurs since $\sigma_D(0,0,0)=\sqrt{3} > \sigma^{fwd}_T(0,0)=1$. The first yield point is the same as in the E-T-D model, and the transformation strain evolves according to (\ref{eq37}) during the T stage. A second yield instant at which damage is triggered exists also in this case, but now, since the transformation yield stress decreases faster than the damage stress with increasing damage, the phase transformation evolves together with damage. During the TD stage, the transformation strain continues to evolve according to (\ref{eq37}) due to the choice $s=2$, whereas damage is given by  
\begin{equation} \label{eq39}
    \alpha=1-\frac{E_0 w_1} {\splitfrac{ ( \tau_{M_0} e^{tr} + \frac{1}{2}h_0 e^{tr^2} + R_0 \overline{e^{tr}})2 E_0 }{+ (\tau_{M_0} + h_0 e^{tr}  + R_0 )^2} }.
\end{equation}
The material response is illustrated in Figure \ref{fig3}, where the coupled evolution of damage and transformation strain in the third stage can be observed. 
\subsubsection{Unloading paths}
\begin{figure*}[t!]
    \centering
    \includegraphics[width=\textwidth]{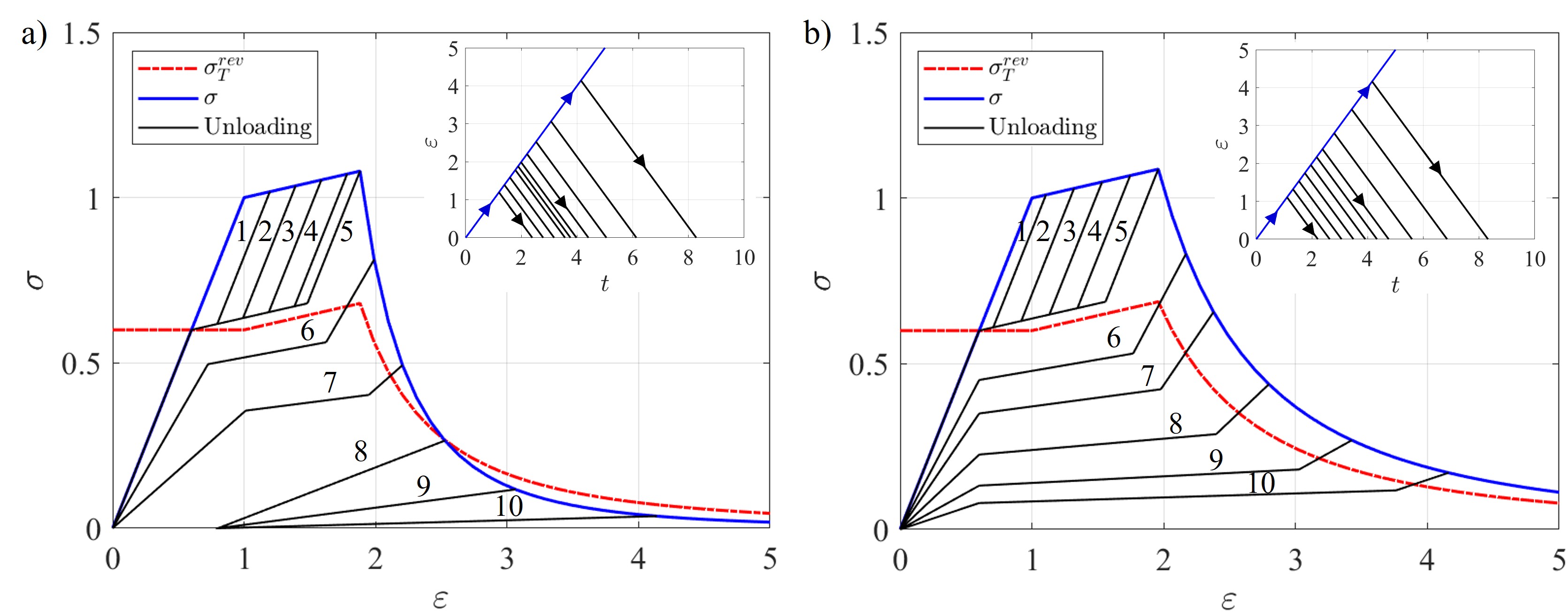}
    \caption{Possible loading and unloading paths for the a) E-T-D model and b) E-T-TD model. The stress reached during loading for different final strains and the corresponding unloading curves are represented in blue and black, respectively. The value of the reverse transformation stress for the different paths is plotted in red. The figures in the upper right corners depict the evolution of the strain during loading (blue) and possible unloading (black).}
    \label{fig4}
\end{figure*}

After analyzing the response of the E-T-D and E-T-TD models under a monotonically increasing strain, it is interesting to discuss what happens if the loading is followed by an unloading path. For both models, if unloading occurs within the transformation stage before the onset of damage, the hysteretic stress-strain curve resembles the well-known flag-shaped response of SMAs, as shown in Figures \ref{fig4}a and \ref{fig4}b for paths 1-5. When unloading occurs within the third stage, different behaviors may be observed for the two models.

In the E-T-D model, if the current stress at the onset of unloading is greater than the reverse transformation stress $\sigma^{rev}_T (e^{tr}, \alpha)$, then unloading consists of a first elastic unloading stage, followed by reverse phase transformation and elastic unloading to zero (see paths 6, 7 in Figure \ref{fig4}a). During the elastic unloading stages, the stress decreases according to the stress-strain relation, whereas the damage and the transformation strain remain constant. During the reverse phase transformation stage, the transformation strain decreases to zero. Moreover, both the elastic stiffness and the slope of the reverse transformation path decrease with the increase of the final damage reached at the end of loading. Conversely, if the stress at the onset of unloading is lower than the reverse transformation stress $\sigma^{rev}_T (e^{tr}, \alpha)$, an elastic unloading occurs, resulting in a residual transformation strain, as in paths 8-10 in Figure \ref{fig4}a. 

In the E-T-TD model, during the third stage the stress at the onset of unloading is equal to the forward transformation stress $\sigma^{fwd}_T (e^{tr}, \alpha)$, hence it is always larger than the reverse transformation stress $\sigma^{rev}_T (e^{tr}, \alpha)$. Consequently, the unloading path always consists of an initial elastic unloading stage, followed by reverse phase transformation with transformation strain recovery, and a final elastic unloading stage to zero, as depicted in Figure \ref{fig4}b (paths 6-10).
%
\subsection{Localization response in a 1D tensile test}
\label{sec4-3}
In this section, we analyze the response of a 1D bar in a tensile test, accounting for the possible spatial localization of the damage and strain variables. The homogeneous response of Section \ref{sec4-1} is possible but is not necessarily the only solution; 
moreover, it may satisfy the first-order but not the second-order local stability condition \citep{Pham2011}. In gradient damage models without plasticity, the homogeneous response is unique and stable if and only if the length $L$ of the bar is sufficiently small compared to the internal length $l$ \citep{Pham2011}. When the length of the bar is large enough with respect to $l$ and for $E(\alpha)$ and $w(\alpha)$ as in \eqref{eq33}, the homogeneous response becomes unstable right at the onset of damage, leading to a non-homogeneous response characterized by damage localization over a time-dependent region of the bar. In our gradient damage model coupling phase transformation with damage, we do not attempt to study second-order stability of the homogeneous solution nor to derive a non-homogeneous solution analytically. Instead, we proceed numerically and we provide numerical evidence of the loss of stability of the homogeneous solution at the onset of damage and of the subsequent localization behavior.
\subsubsection{Numerical approach}
\label{sec_num_approach}
We consider a bar of length $L=1$ and unit cross-sectional area, initially undeformed, not transformed, and undamaged. The end $x=0$ is fixed, while a monotonically increasing displacement $u(L)= U =t$ is applied at the opposite end. We choose $l=0.15$ and adopt as for the homogeneous response $E_0=1$ and $\varepsilon_L \to \infty$.
We consider both the E-T-D and E-T-TD models, using the material parameters in Table \ref{tab2}. 

The time interval $[0,T]$ is discretized in 100 uniform time steps. 
Upon time discretization, the accumulated transformation strain at time $t_i$ is
\begin{equation} \label {eq40}
    \overline{e^{tr}}_i = \overline{e^{tr}}_{i-1} + \lvert e^{tr}_{i} - e^{tr}_{i-1} \rvert.
\end{equation}
For space discretization, we adopt a linear ansatz for the displacement and damage variables and a piecewise constant ansatz for strain and transformation strain. We discretize the bar in 200 linear elements with uniform spacing $h_e=1/200$, which allows for a good approximation of the damage profile.

At the generic time $t_i$ we solve the following minimization problem
\begin{equation}  \label{eq41} 
\min_{u, e^{tr}, \alpha} \{ \mathcal{W}_i (u, e^{tr}, \alpha), \alpha> \alpha_{i-1}, u(0)=0, u(L)=t_i \}
\end{equation}
where the total energy of the bar is rewritten as
\begin{equation}  \label{eq42}
\begin{split}
&\mathcal{W}_i (u, e^{tr}, \alpha) =\\& \int_0^L \biggl( \frac {1}{2} E(\alpha(x)) (u'(x) - e^{tr}(x))^{2} \\&+ \tau_{M}(\alpha(x))\lvert e^{tr}(x)\rvert + \frac{1}{2} h(\alpha(x)) \lvert e^{tr}(x)\rvert^{2} \\
& + R(\alpha(x)) ( \lvert  e^{tr} - e^{tr}_{i-1} \rvert + \overline{e^{tr}}_{i-1}(x)) +  w(\alpha(x)) \\&+ w_{1}l^{2}(\alpha'(x))^{2}\biggr) dx.
\end{split}
\end{equation}
Since (\ref{eq42}) is globally non-convex but is convex with respect to each state variable, for the numerical solution of the governing equations we adopt a staggered algorithm resembling the standard alternate minimization algorithm \citep{Bourdin2000, Bourdin2007, Bourdin2008}, as described in Table \ref{tab3}. First, the energy is minimized with respect to displacement and transformation strain. The displacement is determined by solving a standard linear elastic problem with fixed transformation strain. The transformation strain is then obtained applying a return-mapping algorithm, as described in Table \ref{tab4}. To present the algorithm in a general form, in Tables \ref{tab3} and \ref{tab4} we consider a finite $\varepsilon_{L}$ although in the following examples $\varepsilon_{L} \to \infty$ to simplify the analysis. The role of $\varepsilon_{L}$ will be discussed in Section \ref{sec5-2} when considering cyclic loading conditions. For fixed $u$ and $e^{tr}$, we solve a second minimization with respect to damage. To enforce the damage irreversibility condition, we adopt the penalization method in \cite{Gerasimov2019}. The idea of penalization is to add to the energy functional (\ref{eq42}) the penalty term
\begin{equation}  \label{eq43}
P_{\gamma}(\alpha) = \frac{\gamma}{2} \int_0^L \langle \alpha - \alpha_{i-1}\rangle_{-} ^2 dx, \ \ \ \gamma \gg 1,
\end{equation}
with $\langle y\rangle_{-}=\text{min}(0,y)$. In this way, the originally constrained minimization problem is reduced to the following unconstrained minimization problem
\begin{equation}  \label{eq44}
\min_{u, e^{tr}, \alpha} \{ \mathcal{W}_i (u, e^{tr}, \alpha) + P_{\gamma}(\alpha) , u(0)=0, u(L)=t_i \}.
\end{equation}
For the penalty parameter $\gamma$, we adopt the lower bound proposed by \cite{Gerasimov2019}:
\begin{equation}  \label{eq45}
\gamma= \frac{G_c}{l} \frac{27}{64 \ TOL^2_{ir}},
\end{equation}
with $TOL_{ir} = 0.01$. In (\ref{eq45}), $G_c$ represents the fracture toughness of the material, which is related to the specific fracture energy through 
\begin{equation}  \label{eq46}
G_c=  \frac{8}{3} w_1 l
\end{equation}
for the chosen constitutive function $w(\alpha)=\alpha$.

Note that the equations solving the minimization problems with respect to $(u, e^{tr})$ and $\alpha$ are both non-linear due to the phase transformation and the Macaulay bracket term. Therefore, the Newton-Raphson procedure is used to iteratively compute $u^j$, $e^{tr^j}$, and $\alpha^j$, using $TOL_{NR}=10^{-9}$ as tolerance. The sequence of the two minimization steps is iteratively solved until reaching appropriate tolerances $TOL_{Stag,u}$ and $TOL_{Stag,e^{tr}}$ for the displacement residual and the relative transformation strain variation, respectively, as detailed in Table \ref{tab3}. For simplicity, in this work we take $TOL_{Stag,u}=TOL_{Stag,e^{tr}}=10^{-6}$.
\begin{table*} [t]
\caption{Staggered iterative solution process at a fixed load step $i$.}
\begin{tabularx}{\textwidth} {l}
\hline
\\ \underline{\textbf{Input:}} solution ($u_{i-1}, e^{tr}_{i-1}, \alpha_{i-1}$) from time step $i-1$, and boundary conditions \\
\ \ \ \ \ Initialization, $j=0$:
$u^0_i = u_{i-1}$, $e^{tr^0}_i = e^{tr}_{i-1}$, $\alpha^0_i = \alpha_{i-1}$ \\
\ \ \ \ \ Staggered iteration $j$: \\
\ \ \ \ \ \ \ \ \textbf{do} \\
\ \ \ \ \ \ \ \ \ \ \ \ \ $j=j+1$ \\
\ \ \ \ \ \ \ \ \ \ \ \ \ solve $(u^j_i, e^{tr^j}_i) = \underset{u, e^{tr}}{\mathrm{argmin}} (\mathcal{W}_i (u, e^{tr}, \alpha^{j-1}_{i}) + P_{\gamma}(\alpha^{j-1}_{i})), \ \ \ u(0)=0, u(L)=t_i$  \\
\ \ \ \ \ \ \ \ \ \ \ \ \ (Newton-Raphson) \hspace{0.5cm} See Table \ref{tab4} \\
\ \ \ \ \ \ \ \ \ \ \ \ \ solve $\alpha^j_i = \underset{\alpha}{\mathrm{argmin}} (\mathcal{W}_i (u^j_i, e^{tr^j}_i, \alpha) + P_{\gamma}(\alpha))$ \ \ \ (Newton-Raphson) \\
\ \ \ \ \ \ \ \ \ \ \ \ \ compute the displacement residual:
$R_u = \lvert \lvert \frac{\partial \mathcal{W}_i}{\partial u} (u^j_i, e^{tr^j}_i, \alpha^j_i)\rvert \rvert_2$ \\
\ \ \ \ \ \ \ \ \ \ \ \ \ compute $\sigma (u^j_i, e^{tr^j}_i, \alpha^j_i)$ and $X (\sigma, e^{tr^j}_i, \alpha^j_i)$ \\
\ \ \ \ \ \ \ \ \ \ \ \ \ evaluate the transformation yield function $f_T = \lvert X (\sigma, e^{tr^j}_i, \alpha^j_i) \rvert - R(\alpha^j_i) $\\ \\
\ \ \ \ \ \ \ \ \ \ \ \ \ \textbf{if} $f_T \leq 0$ \textbf{then} \\
\ \ \ \ \ \ \ \ \ \ \ \ \ \ \ \ \ \ $e^{tr^{j,upd}}_i=e^{tr^j}_i$ \\
\ \ \ \ \ \ \ \ \ \ \ \ \ \textbf{else} \\
\ \ \ \ \ \ \ \ \ \ \ \ \ \ \ \ \ \ $e^{tr^{j,upd}}_i= \frac{E(\alpha^j_i) \varepsilon^j_i - \tau_M(\alpha^j_i) - R(\alpha^j_i)}{E(\alpha^j_i)+ h(\alpha^j_i)}$\\ \\
\ \ \ \ \ \ \ \ \ \ \ \ \ \ \ \ \ \ \textbf{if} $e^{tr^{j,upd}}_i \geq \varepsilon_L$ \textbf{then}\\
\ \ \ \ \ \ \ \ \ \ \ \ \ \ \ \ \ \ \ \ \ \ \ $e^{tr^{j,upd}}_i = \varepsilon_L$ \\
\ \ \ \ \ \ \ \ \ \ \ \ \ \ \ \ \ \ \textbf{end if} \\
\ \ \ \ \ \ \ \ \ \ \ \ \ \textbf{end if} \\ 
\ \ \ \ \ \ \ \ \ \ \ \ \ $e^{tr}_{err} = \lvert \lvert e^{tr^{j,upd}}_i - e^{tr^j}_i \rvert \rvert_2$ \\ \\
\ \ \ \ \ \ \ \ \textbf{while} ($R_u \geq TOL_{Stag, u}$) or ( $e^{tr}_{err} \geq TOL_{Stag, e^{tr}}$) \\
\underline{\textbf{Output:}} solution ($u_{i}, e^{tr}_{i}, \alpha_{i}$), with $u_{i}=u^j_i, e^{tr}_{i}=e^{tr^{j,upd}}_i$, and $\alpha_{i}=\alpha^j_i$.\\ \\
\hline
\end{tabularx}
\label{tab3}
\end{table*}
\begin{table*} [h]
\caption{Minimization problem with respect to $(u, e^{tr})$ at fixed time step $i$ and staggered iteration $j$.}
\begin{tabularx}{\textwidth} {l}
\hline
\\ \underline{\textbf{Input:}} solution ($u^{j-1}_{i}, e^{tr^{j-1}}_{i}, \alpha^{j-1}_{i}$) from staggered iteration $j-1$, and boundary conditions \\
\ \ \ \ \ Initialization, $k=0$: $e^{tr^0} = e^{tr^{j-1}}_{i}$ \\
\ \ \ \ \ \ \ \ \textbf{do} \\
\ \ \ \ \ \ \ \ \ \ \ \ \ $k=k+1$ \\
\ \ \ \ \ \ \ \ \ \ \ \ \ solve $u^k= \underset{u}{\mathrm{argmin}} (\mathcal{W}_i (u, e^{tr^{k-1}}, \alpha^{j-1}_{i}) + P_{\gamma}(\alpha^{j-1}_{i})), \ \ \ u(0)=0, u(L)=t_i$ \\
\ \ \ \ \ \ \ \ \ \ \ \ \ (Newton-Raphson)\\
\ \ \ \ \ \ \ \ \ \ \ \ \ compute $\sigma (u^k, e^{tr^{k-1}}, \alpha^{j-1}_i)$ and $X (\sigma, e^{tr^{k-1}}, \alpha^{j-1}_i)$ \\
\ \ \ \ \ \ \ \ \ \ \ \ \ evaluate the transformation yield function $f_T = \lvert X (\sigma, e^{tr^{k-1}}, \alpha^{j-1}_i) \rvert - R(\alpha^{j-1}_i)$ \\ \\
\ \ \ \ \ \ \ \ \ \ \ \ \ \textbf{if} $f_T \leq 0$ \textbf{then} \\
\ \ \ \ \ \ \ \ \ \ \ \ \ \ \ \ \ \ $e^{tr^k}=e^{tr^{k-1}}$ \\
\ \ \ \ \ \ \ \ \ \ \ \ \ \textbf{else} \\
\ \ \ \ \ \ \ \ \ \ \ \ \ \ \ \ \ \ $e^{tr^k}= \frac{E(\alpha^{j-1}_i) \varepsilon^k - \tau_M(\alpha^{j-1}_i) - R(\alpha^{j-1}_i)}{E(\alpha^{j-1}_i)+ h(\alpha^{j-1}_i)}$\\ \\
\ \ \ \ \ \ \ \ \ \ \ \ \ \ \ \ \ \ \textbf{if} $e^{tr^k} \geq \varepsilon_L$ \textbf{then}\\
\ \ \ \ \ \ \ \ \ \ \ \ \ \ \ \ \ \ \ \ \ \ \ $e^{tr^k} =\varepsilon_L$ \\
\ \ \ \ \ \ \ \ \ \ \ \ \ \ \ \ \ \ \textbf{end if} \\
\ \ \ \ \ \ \ \ \ \ \ \ \ \textbf{end if} \\
\ \ \ \ \ \ \ \ \ \ \ \ \ $e^{tr}_{err} = \lvert \lvert e^{tr^k} - e^{tr^{k-1}} \rvert \rvert_2$ \\ \\
\ \ \ \ \ \ \ \ \textbf{while} $e^{tr}_{err} \geq 10^{-6}$ \\
\underline{\textbf{Output:}} solution ($u^j_i, e^{tr^j}_i$) of Table \ref{tab3}, with $u^j_i=u^k$, $e^{tr^j}_i=e^{tr^k}$.\\  \\
\hline
\end{tabularx}
\label{tab4}
\end{table*}
\subsubsection{E-T-D model}
\begin{figure*}[t!]
    \centering
    \includegraphics[width=\textwidth]{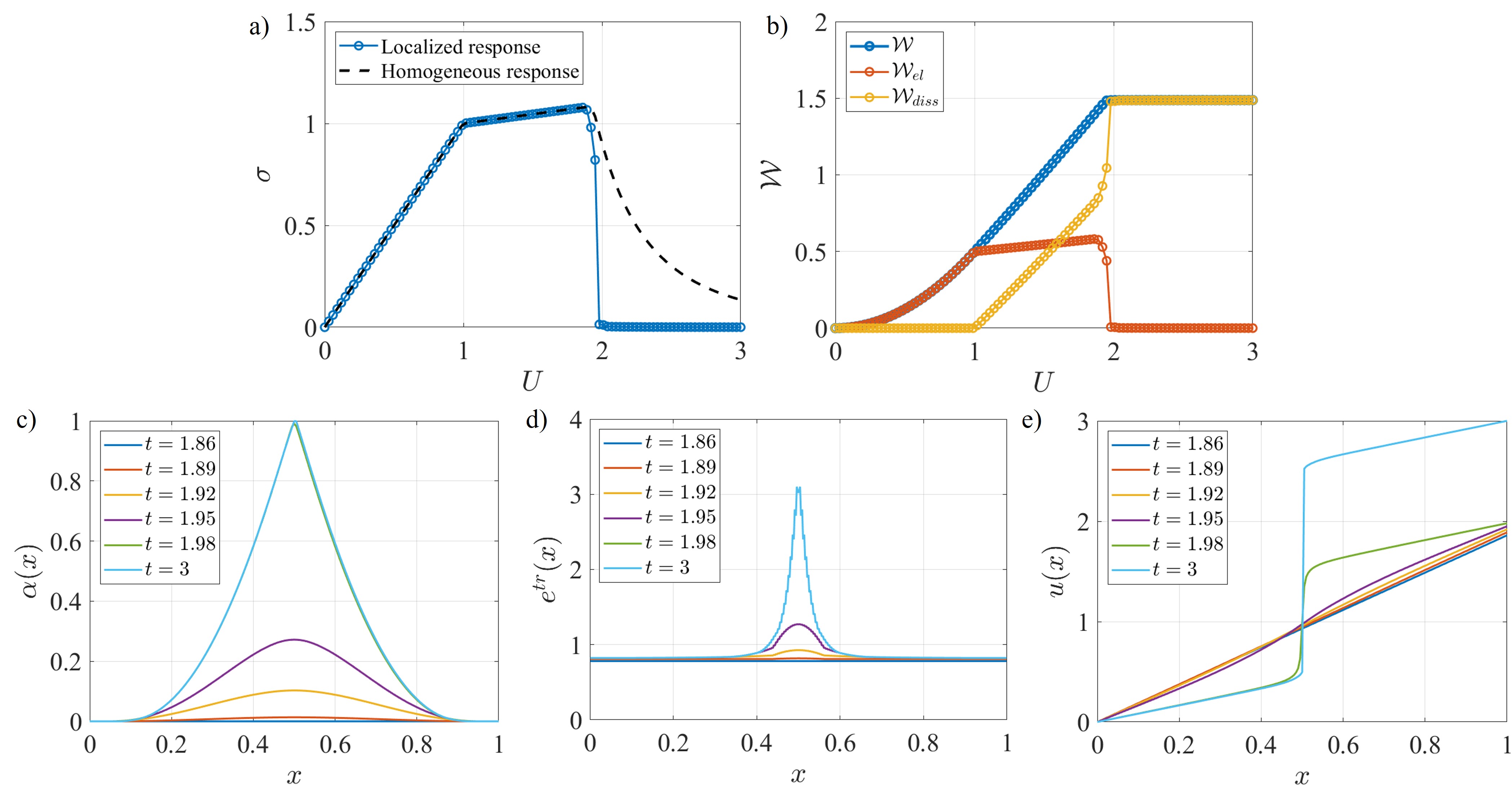}
    \caption{Localized response for the E-T-D model with $l=0.15$: a) stress-displacement diagram; b) energy diagrams; c) damage, d) transformation strain, and e) displacement profiles along the bar at different loading steps.}
    \label{fig5}
\end{figure*}

Up to the onset of damage, we observe the homogeneous response described in Section \ref{sec4-1-1}, depicted as a dashed line in Figure \ref{fig5}a. When the displacement reaches the second yield point, damage is triggered. Unlike in the homogeneous response, at the structural level damage evolution is coupled with transformation strain localization and the transformation strain continues to evolve.
For $l=0.15$, the stress initially follows a softening branch (Figure \ref{fig5}a), indicative of cohesive fracture behavior. After a few time increments, the stress suddenly vanishes, with the damage variable reaching value 1 at the centerpoint of the bar (Figure \ref{fig5}c). This behavior could be attributed either to the system approaching the stability limit of the softening response for this specific $L/l$ ratio, or to a spurious numerical artifact. Clarification of this aspect would require a second-order stability analysis, which is beyond the scope of this work.
The damage localizes, suddenly reaching its maximum value at the singular point at the center of the bar, leading to a gradient singularity (Figure \ref{fig5}c). The transformation strain localizes as well at the center of the bar and keeps a constant value outside the localization region, equal to the transformation strain reached at the second yield point (Figure \ref{fig5}d). The displacement profile in Figure \ref{fig5}e shows a jump at the center of the localization area which is related to the transformation strain singularity, indicating the occurrence of a crack. The total strain, related to the non-vanishing slope of the displacement profile outside the localization region, compensates the accumulated transformation strain. 
The energy plot in Figure \ref{fig5}b displays the evolution of the total energy of the bar $\mathcal{W} (u, e^{tr},  \overline{e^{tr}}, \alpha)$ (\ref{eq24}), given by the sum of the elastic energy and the total dissipated energy, respectively given by
 \begin{equation}  
 \mathcal{W}_{el} (u, e^{tr}, \alpha) = 
 \int_{0}^{L}  \frac{1}{2} E(\alpha) (\varepsilon(u)-e^{tr})^2 dx,
 \end{equation}
 \begin{equation} 
 \begin{split}
 & \mathcal{W}_{diss}(e^{tr}, \overline{e^{tr}}, \alpha) = \\&
 \int_{0}^{L}  \biggl[\tau_{M}(\alpha)\lvert e^{tr}\rvert + \frac{1}{2} h(\alpha) \lvert e^{tr}\rvert^{2} + R(\alpha) \overline{e^{tr}} \\& + w(\alpha) + w_{1}l^{2}(\alpha')^{2}\biggr] dx.
 \end{split}
 \end{equation}
The total dissipated energy, due to both phase transformation and damage, initially shows a smooth increase during the stress softening branch, followed by an instantaneous energy dissipation when the stress goes to zero. 

\begin{figure*}[tp!]
    \centering
    \includegraphics[width=\textwidth]{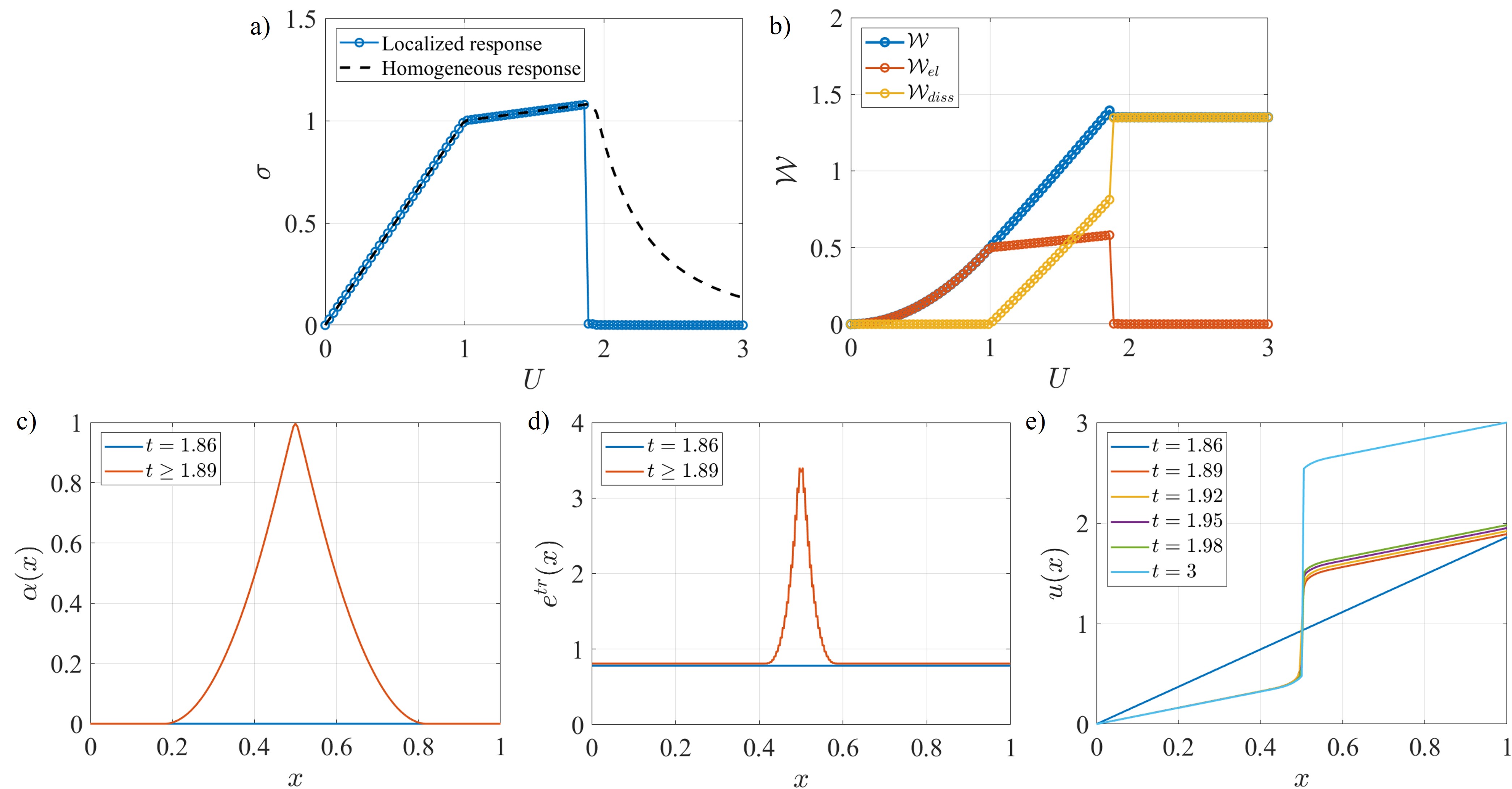}
    \caption{Localized response for the E-T-D model with $l=0.12$: a) stress-displacement diagram; b) energy diagrams; c) damage, d) transformation strain, and e) displacement profiles along the bar at different loading steps.}
    \label{fig6}
\end{figure*}

The numerical response is strongly influenced by the internal length $l$. For larger values of $l$, a larger part of the stress softening branch can be captured, recovering the homogeneous response for $l \geq L$. Decreasing $l$, the softening branch is replaced by the abrupt formation of a crack, as visible in Figure \ref{fig6}a for $l=0.12$. In this case, the stress drops immediately to zero, with damage reaching value 1 at the center of the bar (Figure \ref{fig6}c) and instantaneous energy dissipation (Figure \ref{fig6}b). The displacement profile in Figure \ref{fig6}e presents a jump discontinuity with localized transformation strain (Figure \ref{fig6}d) and non-vanishing total strain outside the localization region.
\subsubsection{E-T-TD model}
\begin{figure*}[htbp]
    \centering
    \includegraphics[width=\textwidth]{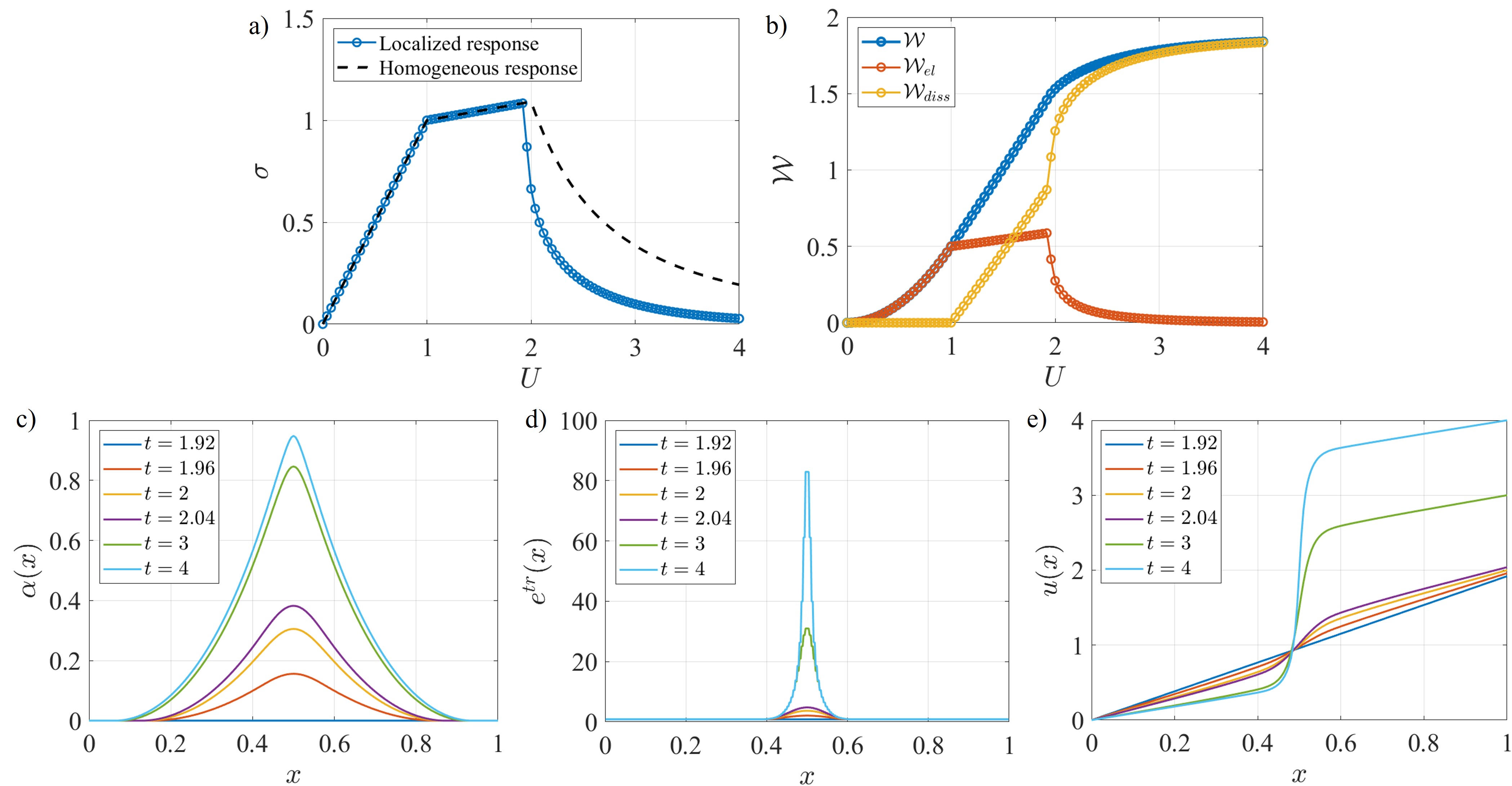}
    \caption{Localized response for the E-T-TD model with $l=0.15$: a) stress-displacement diagram; b) energy diagrams; c) damage, d) transformation strain, and e) displacement profiles along the bar at different loading steps.}
    \label{fig7}
\end{figure*}

Also with this model, we observe the homogeneous response up to the onset of damage, depicted as a dashed line in Figure \ref{fig7}a.
At the second yield point, a cohesive fracture appears at the center of the bar, characterized by a discontinuity in displacement (Figure \ref{fig7}e) with non-vanishing stress. For $l=0.15$, the stress immediately decreases to a value that is about 80\% of the transformation yield stress. Then, it reduces asymptotically towards zero for an infinite opening of the crack (Figure \ref{fig7}a), resembling a Barenblatt-like cohesive response. During the fracturing TD stage, the damage and the transformation strain simultaneously evolve, as visible in Figure \ref{fig7}c and Figure \ref{fig7}d. The energy profiles in Figure \ref{fig7}b show a jump in the dissipated energy which then increases smoothly with loading, due to the evolution of damage and transformation strain. 

Also in this case, the internal length plays a fundamental role in the numerical response. Increasing $l$ such that $l \geq L$, the homogeneous response is recovered. Decreasing $l$, the jump in the stress at the onset of damage becomes more abrupt.

Note that, since in these examples the material parameter $\varepsilon_L$ is assumed to be infinitely large, the localized behavior of the material under monotonic loading resembles that of standard elastoplastic materials with stress hardening \citep{Alessi2018}. The role of the parameter $\varepsilon_L$ will be addressed in Section \ref{sec5-2} when considering fatigue effects.

\section{Fatigue response}
\label{sec5}
An interesting feature of the present model is its capability to describe fatigue effects leading to the evolution of damage under cyclic loads. This property stems from the dependence of the energy density on the accumulated transformation strain and from its coupling with damage. Under cyclic loads, the accumulated transformation strain increases monotonically, leading to a progressive reduction in the damage yield stress and to the consequent damage evolution due to fatigue. Therefore, fatigue effects can be modeled just exploiting the structure of the model, preserving its variational nature and without the need to introduce additional variables. 

Consider a bar of length $L=1$, subjected to a cyclic displacement applied at its end $x=L$ ranging between $0$ and $U_0$, corresponding to a homogeneous or average strain ranging between $0$ and $\varepsilon_0 = U_0/L$. We assume cyclic loading in pure tension since this is the most interesting condition for fatigue-relevant SMA cardiovascular applications. 
Denoting as $\varepsilon_T$ and $\varepsilon_D$ the total strain at which forward phase transformation and damage are triggered, respectively, we assume that $ \varepsilon_T < \varepsilon_0 < \varepsilon_D$, such that the bar enters the phase transformation stage before damaging and no damage is attained during the first half-cycle. 
\subsection{Homogeneous response}
\begin{figure}[t]
    \centering
    \includegraphics[width=\columnwidth]{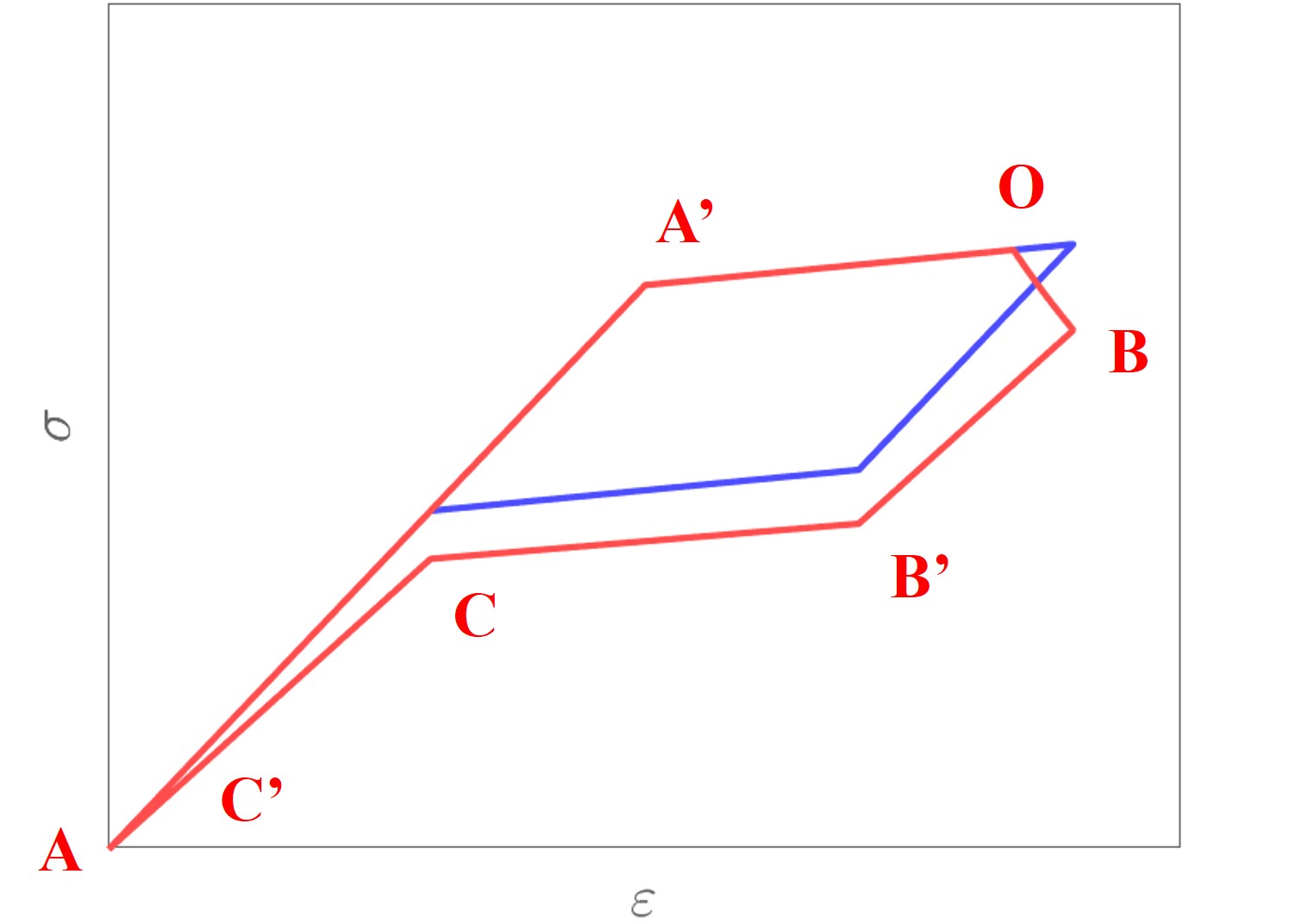}
    \caption{Exemplary stress-strain relationship during a fatigue cycle without damage (in blue) and during the first cycle characterized by homogeneous damage evolution (in red).}
    \label{fig8}
\end{figure}

The homogeneous response of the bar (or the response of a material point) is described by the exemplary stress-strain plot in Figure \ref{fig8}. 
Up to the onset of damage, the stress-strain behavior has the typical flag shape which is characteristic of SMAs, as illustrated by the blue curve in Figure \ref{fig8}. With repeated loading cycles, $\overline{e^{tr}}$ accumulates during each forward and reverse phase transformation stage. This accumulation progressively reduces the damage yield stress, ultimately leading to damage evolution after a certain number of cycles. Suppose that the red curve in Figure \ref{fig8} represents the cycle at which damage is triggered for the first time. 
The initial AA' stage is elastic, with zero damage, zero transformation strain, and fixed accumulated transformation strain.
During the A'B path, the material undergoes forward phase transformation. When the damage yield stress is reached (point O), damage is triggered, and from that point a coupled evolution of damage and transformation strain takes place. During the OB stage, the transformation strain and the accumulated transformation strain can be
obtained by equating the stress  with $\sigma_T(e^{tr}, \alpha)$, and imposing $\sigma_T(e^{tr}, \alpha)=\sigma_D(e^{tr}, \overline {e^{tr}}, \alpha)$, resulting in
\begin{equation}  \label{eq47}
e^{tr} (\alpha)= \frac{ (1-\alpha)^2 E_0  \varepsilon - (1-\alpha)^s (\tau_{M_0} + R_0)}{(1-\alpha)^2 E_0 + (1-\alpha)^s h_0} ,
\end{equation}
\begin{equation}  \label{eq48}
\overline{e^{tr}} (\alpha)= \frac{ \begin{split} &(1-\alpha)^3 E_0 w_1  \\& - (1-\alpha)^{2s} (\tau_{M_0} + h_0 e^{tr}(\alpha) + R_0)^2 \\& -  (1-\alpha)^{s+2} s E_0 (\tau_{M_0} e^{tr}(\alpha) + \frac{1}{2} h_0 e^{tr^2}(\alpha)) \end{split}} 
{ (1-\alpha)^{s+2} s E_0 R_0 }.
\end{equation}
To solve for damage, the relation between $e^{tr}$ and $\overline {e^{tr}}$ is used as the third equation, exploiting the known value of accumulated transformation strain at the beginning of the cycle $\overline{e^{tr}}_A$:
\begin{equation}  \label{eq49}
\overline{e^{tr}}(\alpha) = \overline{e^{tr}}_A +  e^{tr}(\alpha).
\end{equation}
Substituting (\ref{eq47}) and (\ref{eq48}) into (\ref{eq49}), a single equation in the damage variable is obtained, whose solution provides the strain-driven evolution of $\alpha$. Subsequently, $e^{tr}$ and $\overline {e^{tr}}$ are updated according to (\ref{eq47}) and (\ref{eq48}), respectively.
During the elastic unloading stage BB', damage and transformation strain remain constant, and the slope $E(\alpha)$ is reduced due to the previous damage evolution. During the subsequent reverse transformation path B'C, $e^{tr}$ decreases to zero while $\overline {e^{tr}}$ increases, leading to a decrease of the forward and reverse transformation yield stresses and to an increase of the damage yield stress according to (\ref{eq35}) and (\ref{eq36}). Consequently, the damage yielding condition is never attained during unloading and damage remains constant, equal to the value reached at point B. The final elastic unloading stage CC' occurs at zero transformation strain and fixed damage and accumulated transformation strain.

The homogeneous response under fatigue cycles is obtained numerically as schematized in Table \ref{tab5}. The strain resulting from the cyclic boundary conditions is adopted as control variable, and the time interval is discretized in equally-spaced time steps. At each time step, an elastic prediction is computed for all the state variables. Assuming a fixed damage, the transformation yield criterion is first checked both during loading and unloading, implementing the standard return-mapping algorithm to eventually update $e^{tr}$, $\overline{e^{tr}}$ and the stress. During the loading path, the damage yield criterion is then verified as a nested condition, updating the state variables if needed. 
Note that the return-mapping algorithm in Table \ref{tab5} is written for a finite $\varepsilon_L$, above which the material behaves elastically in the fully martensitic phase.
\begin{table*} [t!]
\caption{Numerical computation of the homogeneous response under fatigue loading considering cycle $i$ and time step $j$. The initialization values $\varepsilon^{i-1}_{end}$, $e^{{tr}^{i-1}}_{end}$, $\overline{e^{{tr}^{i-1}}_{end}}$, $\sigma^{i-1}_{end}$, and $\alpha^{i-1}_{end}$ are, respectively, the strain, the transformation strain, the accumulated transformation strain, the stress, and the damage computed at the last time step of the previous cycle $i-1$.}
\begin{tabularx}{\textwidth} {l}
 \hline
 \underline{\textbf{Loading half-cycle:}} \\
\ \ \ \ \ Initialization, $j=0$: $\varepsilon_0 =  \varepsilon^{i-1}_{end}, e^{tr}_0 = e^{{tr}^{i-1}}_{end}, \overline{e^{tr}_0} = \overline{e^{{tr}^{i-1}}_{end}}, \sigma_0 = \sigma^{i-1}_{end}, \alpha_0 = \alpha^{i-1}_{end}$\\
\ \ \ \ \ Time step $j \geq 1$:\\
\ \ \ \ \ \ \ \ compute an elastic increment: $\varepsilon_j =  \varepsilon_{j-1} + \Delta \varepsilon, e^{tr}_j = e^{tr}_{j-1}, \overline{e^{tr}_j} = \overline{e^{tr}_{j-1}}, \alpha_j = \alpha_{j-1}$\\
\ \ \ \ \ \ \ \ compute the elastic stress $\sigma_j$\\
\ \ \ \ \ \ \ \ \textbf{if}  $\sigma_j \geq \sigma_T^{fwd}(e^{tr}_j, \alpha_j)$ \textbf{then} \\ 
\ \ \ \ \ \ \ \ \ \ \ \ \ projection: 
$e^{tr}_j = \frac{ (1-\alpha_j)^2 E_0  \varepsilon_j - (1-\alpha_j)^s (\tau_{M_0} + R_0)}{(1-\alpha_j)^2 E_0 + (1-\alpha_j)^s h_0}$ \\ \\
\ \ \ \ \ \ \ \ \ \ \ \ \ \textbf{if} $e^{tr}_j < \varepsilon_L$ \textbf{then} \\ 
\ \ \ \ \ \ \ \ \ \ \ \ \ \ \ \ \ \ forward transformation stage:
$\overline{e^{tr}_j} = \overline{e^{tr}_{j-1}} + (e^{tr}_{j} - e^{tr}_{j-1})$, $\sigma_j = \sigma_T^{fwd}(e^{tr}_j, \alpha_j)$  \\ \\
\ \ \ \ \ \ \ \ \ \ \ \ \ \ \ \ \ \ \textbf{if} $\sigma_j \geq \sigma_D(e^{tr}_j, \overline{e^{tr}_j}, \alpha_j)$  \textbf{then} \\ 
\ \ \ \ \ \ \ \ \ \ \ \ \ \ \ \ \ \ \ \ \ \ \ damage evolution: compute $\alpha_j, e^{tr}_{j}$ and $\overline{e^{tr}_j}$ through (\ref{eq47})-(\ref{eq49}), and update the stress\\ 
\ \ \ \ \ \ \ \ \ \ \ \ \ \ \ \ \ \ \textbf{end if}\\ 
\ \ \ \ \ \ \ \ \ \ \ \ \ \textbf{else} \\
\ \ \ \ \ \ \ \ \ \ \ \ \ \ \ \ \ \ elastic loading: 
$e^{tr}_j = \varepsilon_L$, $\overline{e^{tr}_j} = \overline{e^{tr}_{j-1}} + (e^{tr}_{j} - e^{tr}_{j-1})$, $\sigma_j = E_0 (1-\alpha_j)^2 (\varepsilon_j - e^{tr }_j)$\\ \\
\ \ \ \ \ \ \ \ \ \ \ \ \ \ \ \ \ \ \textbf{if} $\sigma_j \geq \sigma_D(e^{tr}_j, \overline{e^{tr}_j}, \alpha_j)$ \textbf{then} \\ 
\ \ \ \ \ \ \ \ \ \ \ \ \ \ \ \ \ \ \ \ \ \ \ damage evolution: 
compute $\alpha_j$ imposing $\sigma_j = \sigma_D(e^{tr}_j, \overline{e^{tr}_j},\alpha)$ and update the stress \\ 
\ \ \ \ \ \ \ \ \ \ \ \ \ \ \ \ \ \ \textbf{end if} \\ 
\ \ \ \ \ \ \ \ \ \ \ \ \ \textbf{end if} \\ 
\ \ \ \ \ \ \ \ \textbf{else} \\
\ \ \ \ \ \ \ \ \ \ \ \ \ elastic stage \\ 
\ \ \ \ \ \ \ \ \textbf{end if} \\ 
\\ \underline{\textbf{Unloading half-cycle:}} \\
\ \ \ \ \ \ \ \ compute an elastic increment: $\varepsilon_j =  \varepsilon_{j-1} - \Delta \varepsilon, e^{tr}_j = e^{tr}_{j-1}, \overline{e^{tr}_j} = \overline{e^{tr}_{j-1}}, \alpha_j = \alpha_{j-1}$\\
\ \ \ \ \ \ \ \ compute the elastic stress $\sigma_j$\\
\ \ \ \ \ \ \ \ \textbf{if}  $ \sigma_T^{rev}(0, \alpha_j) \leq \sigma_j \leq \sigma_T^{rev}(e^{tr}_j, \alpha_j)$ \textbf{then} \\
\ \ \ \ \ \ \ \ \ \ \ \ \ projection: 
$e^{tr}_j = \frac{ (1-\alpha_j)^2 E_0  \varepsilon_j - (1-\alpha_j)^s (\tau_{M_0} - R_0)}{(1-\alpha_j)^2 E_0 + (1-\alpha_j)^s h_0}$, 
$\overline{e^{tr}_j} = \overline{e^{tr}_{j-1}} + \lvert e^{tr}_{j} - e^{tr}_{j-1} \rvert$, $\sigma_j = \sigma_T^{rev}(e^{tr}_j, \alpha_j)$  \\ 
\ \ \ \ \ \ \ \ \textbf{else} \\ 
 \ \ \ \ \ \ \ \ \ \ \ \ \ elastic stage \\
\ \ \ \ \ \ \ \ \textbf{end if}\\ 
\hline
\end{tabularx}
\label{tab5}
\end{table*}

\begin{figure*}[thb!]
    \centering
    \includegraphics[width=0.85\textwidth]{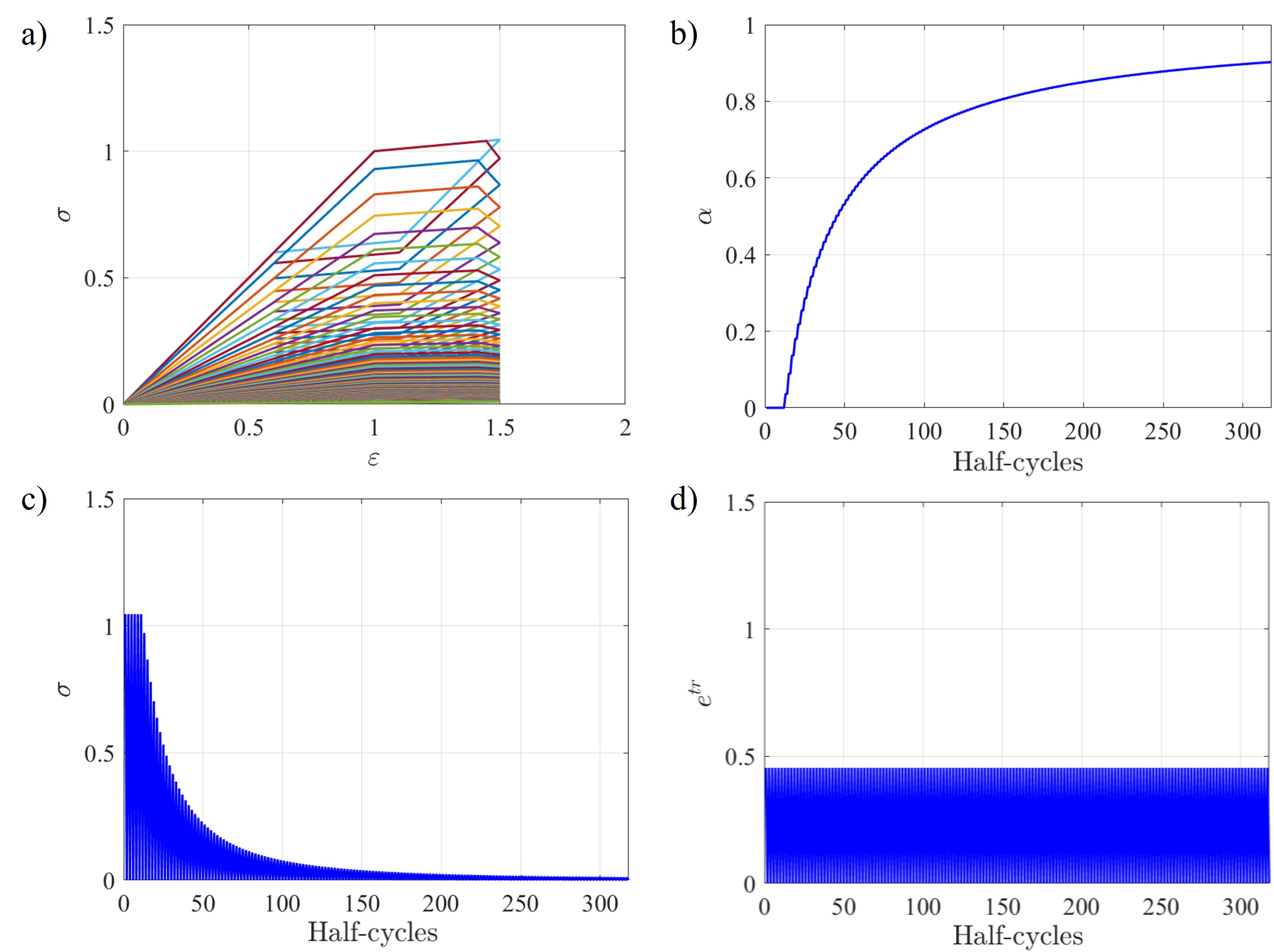}
    \caption{E-T-TD model with $\varepsilon_0=1.5$ and $s=2$: a) stress-strain response under repeated fatigue cycles with progressive damage evolution; b) evolution of damage versus the number of half-cycles; c) stress evolution versus the number of half-cycles; d) transformation strain oscillations versus the number of half-cycles.}
    \label{fig9}
\end{figure*}

Two numerical examples are presented in the following to illustrate fatigue effects, with the material parameters of the E-T-TD model in Table \ref{tab2}, $E_0=1$, and $\varepsilon_L \to \infty$. In the first example, we apply a cyclic strain oscillating between $0$ and $1.5$, with the results reported in Figure \ref{fig9}. The stress-strain response is progressively modified during cyclic loading, as visible in Figure \ref{fig9}a. Specifically, the slopes of both the elastic stages and the phase transformation paths decrease due to damage evolution. Damage starts to evolve during the forward transformation stage in the $7^{\text{th}}$ cycle, when the damage yield criterion is reached, leading to a reduction in the stress. Damage also affects the forward and reverse transformation stresses, with stress plateaus progressively shifting towards lower values. It is interesting to note that no damage occurs during both the elastic unloading and the reverse transformation stages. Indeed, during elastic unloading, the damage yield stress remains constant since $e^{tr}$ and $\overline {e^{tr}}$ are fixed, and the stress decreases below its value. During the reverse transformation, the damage yield stress is initially equal to the forward transformation stress at the end of the loading half-cycle, and it increases due to the decrease of $e^{tr}$; the reverse transformation stress is always lower than the forward transformation stress, so that the damage yield criterion is always satisfied as inequality and damage does not evolve. Figure \ref{fig9}b and Figure \ref{fig9}c show that the damage tends asymptotically to 1, while the stress during subsequent half-cycles oscillates between zero and a peak stress tending asymptotically to zero. Introducing a threshold on the peak stress, a finite fatigue life is obtained, equal in this case to 159 cycles for a threshold of 0.01. The transformation strain oscillates between zero and a maximum constant value (Figure \ref{fig9}d). This is due to the choice of the parameter $s=2$ which leads the damage-dependent terms to cancel out in (\ref{eq47}), thereby removing the dependence of the transformation strain on damage. 

\begin{figure*}[thb!]
    \centering
    \includegraphics[width=0.85\textwidth]{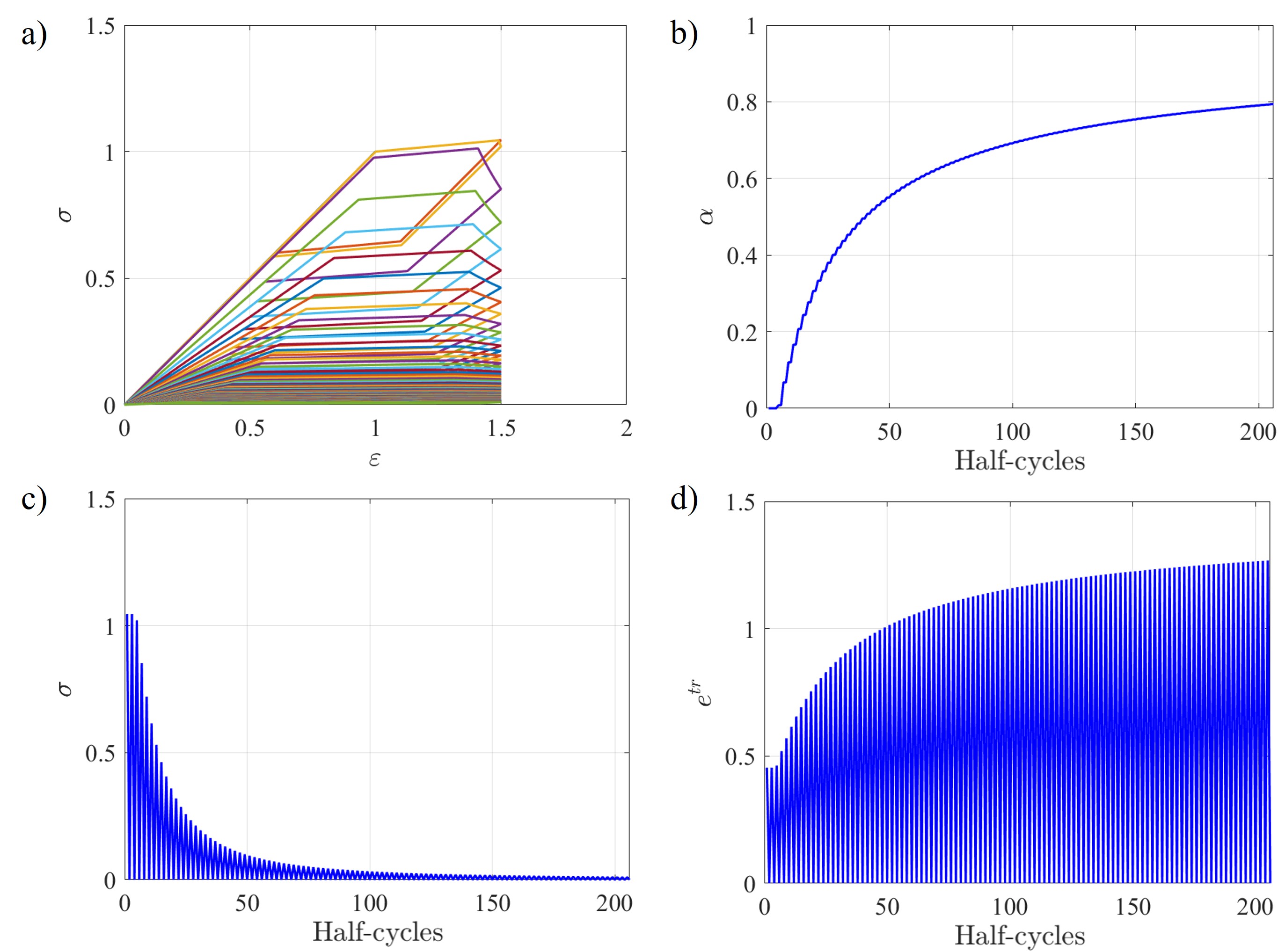}
    \caption{E-T-TD model with $\varepsilon_0=1.5$ and $s=3$: a) stress-strain response under repeated fatigue cycles with progressive damage evolution; b) evolution of damage versus the number of half-cycles; c) stress evolution versus the number of half-cycles; d) transformation strain oscillations versus the number of half-cycles.}
    \label{fig10}
\end{figure*}

In the second example, $s$ is changed to 3, keeping the other parameters and the boundary conditions fixed. In this case, 3 cycles are needed for damage to start evolving. As visible in Figure \ref{fig10}a, the shape of the stress-strain curves progressively changes as previously described. The damage tends asymptotically to 1, with the stress tending asymptotically to zero (Figure \ref{fig10}b and \ref{fig10}c). A finite fatigue life of 103 cycles is computed by taking a threshold on the peak stress of 0.01. In this case, due to the choice $s=3$, the transformation strain at the peak of each cycle progressively increases with damage (Figure \ref{fig10}d).
Analyzing the influence of the softening parameter $s$, damage initiates earlier for $s=3$ compared to $s=2$. Moreover, for $s=3$, the stress decreases more rapidly during the initial damaging cycles, and its variation progressively reduces with increasing number of cycles. The evolution of damage shows a similar behavior, resulting in a lower final damage value for $s=3$ than for  $s=2$, at the same stress threshold.
These simple examples demonstrate how fatigue damage evolution can be easily described by the present model, thanks to the interplay between the accumulation of transformation strain and damage.
\subsection{Localization response in a 1D fatigue tensile test}
\label{sec5-2}
In this section we analyze the response of a 1D bar in a  fatigue tensile test, accounting for the possible spatial localization of the damage and strain variables. The evolution problem is solved numerically as described in Section \ref{sec_num_approach} for the monotonic localized response. The numerical algorithms presented in Tables \ref{tab3} and \ref{tab4} are slightly modified to deal with cyclic boundary conditions and reverse phase transformation during the unloading half-cycles. The time interval during each half-cycle is discretized in 50 uniform time steps. The geometry and the spatial discretization are as described in Section \ref{sec_num_approach}, with $L=1$, $l=0.15$, and uniform element size $h_e=1/200$.

\begin{figure*}[thpb]
    \centering
    \includegraphics[width=\textwidth]{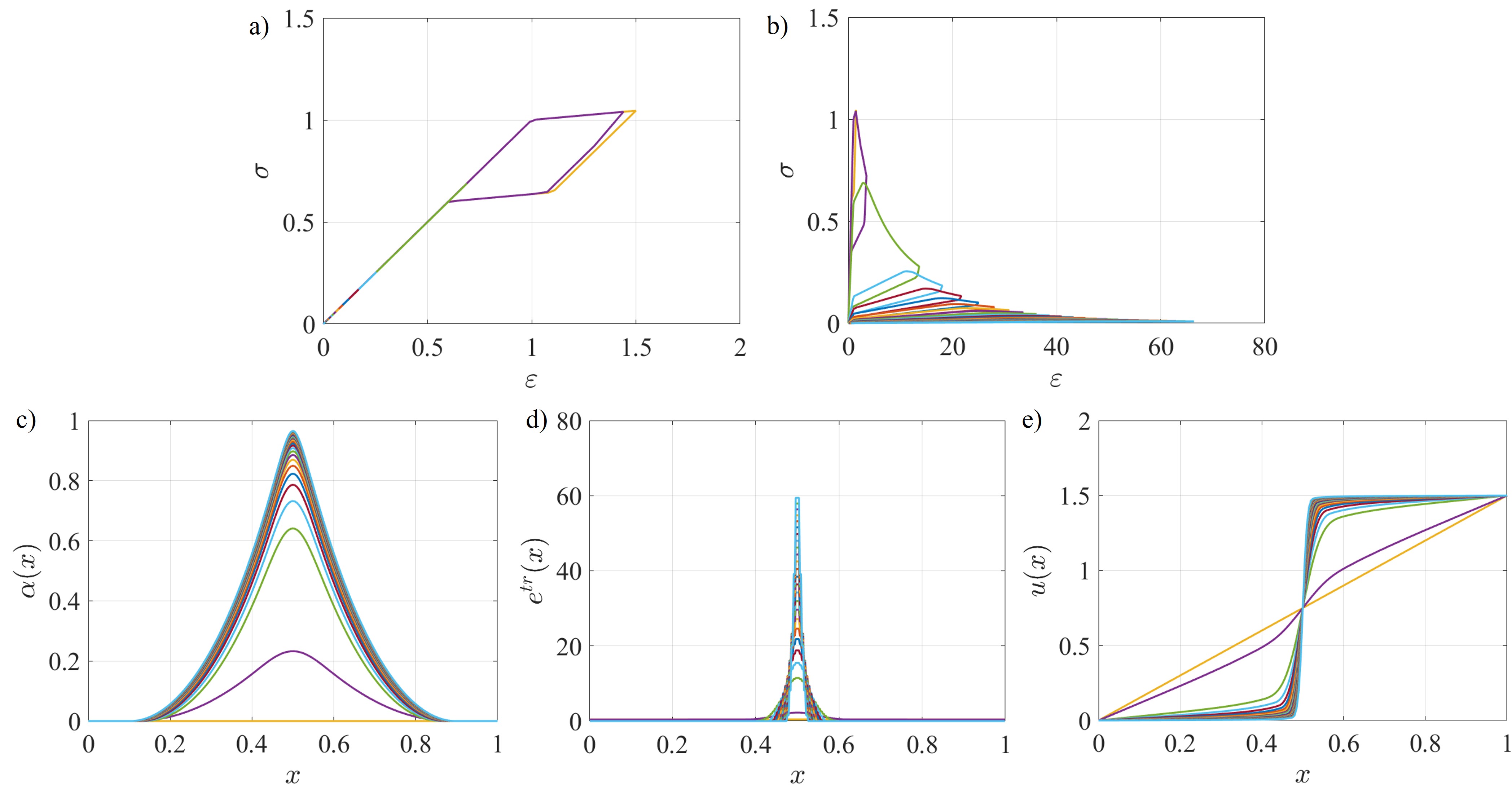}
    \caption{Localized response for the E-T-TD model with $\varepsilon_L \to \infty$ and U ranging between 0 and 1.5: a) local stress-strain curves at the last element under fatigue cycles; b) local stress-strain curves at the central element under fatigue cycles; c) damage profile along the bar at the peak stress of subsequent fatigue cycles; d) transformation strain profile along the bar at the peak stress of subsequent fatigue cycles; e) displacement profile along the bar at the peak stress of subsequent fatigue cycles.}
    \label{fig11}
\end{figure*}

To illustrate fatigue effects, we apply a cyclic displacement $U$ at the end of the bar, oscillating between 0 and 1.5, resulting in a cyclic average strain between 0 and 1.5.
In the first example, we adopt the parameters of the E-T-TD model fixing $s=2$, with $E_0=1$ and $\varepsilon_L \to \infty$. 
The results are reported in Figure \ref{fig11}, showing the local stress-strain response at the element at the end and at the center of the bar, and the damage, transformation strain, and displacement profiles. In this case, damage starts earlier compared to the corresponding homogeneous response ($4^{\text{th}}$ versus $7^{\text{th}}$ cycle). 
As soon as damage starts to evolve, the material is rapidly unloaded. In this specific example, the local stress-strain response evaluated at the element at the end of the bar exhibits just one hysteretic cycle followed by elastic cycles in the austenitic phase (Figure \ref{fig11}a). 
Conversely, at the center of the bar the material shows hysteretic stress-strain cycles progressively shifting towards lower stress values, with decreasing slope of both the elastic and phase transformation paths (Figure \ref{fig11}b). Damage increases during the forward phase transformation stages and its evolution stops during unloading, as in the homogeneous response. 
All local stress-strain cycles after damage nucleation assume the same morphology related to different response stages: elastic loading in austenite, phase transformation, damage trigger with coupled evolution of damage and transformation, elastic unloading, reverse transformation, and elastic unloading to zero. At the center of the bar, the peak strain under cyclic loads increases above the nominal value due to the localization of deformation. 
The stress is uniform in the bar due to equilibrium; it rapidly decreases during the first damaging cycles and then it undergoes small variations during subsequent cyclic loads. 
As soon as damage initiates, the damage and transformation strain localize on a support region centered on the singular point at the center of the bar. 
Figure \ref{fig11}c depicts the damage profiles at the fatigue peaks during subsequent cycles. The damage assumes the profile seen in Section \ref{sec4-3}, with a peak value at the singular point and a discontinuity in the gradient. The maximum damage value increases rapidly during the first damaging cycles and then undergoes smaller variations during subsequent cyclic loads. The discontinuity in the damage gradient is associated to a singularity in the transformation strain, as depicted in Figure \ref{fig11}d. Due to the assumption $\varepsilon_L \to \infty$, the transformation strain localizes and its value at the singular point progressively increases during subsequent cyclic loads, locally tending to infinity. The displacement profile in Figure \ref{fig11}e shows a jump at the center of the localization area which is related to the transformation strain singularity. It initially shows a non-vanishing slope outside the localization region due to the residual total strain; then, it progressively flattens. A finite fatigue life of 27 cycles is computed by introducing a threshold of 0.01 on the peak stress. This result is significantly lower than the fatigue life predicted under the assumption of uniform damage and transformation strain along the bar (159 cycles). In the homogeneous case, the entire bar contributes to resisting fracture, thereby delaying the damaging phenomenon and increasing the fatigue life. However, for a sufficiently long bar the homogeneous solution is unstable and does not reflect the actual failure behavior.

\begin{figure*}[thpb!]
    \centering
    \includegraphics[width=\textwidth]{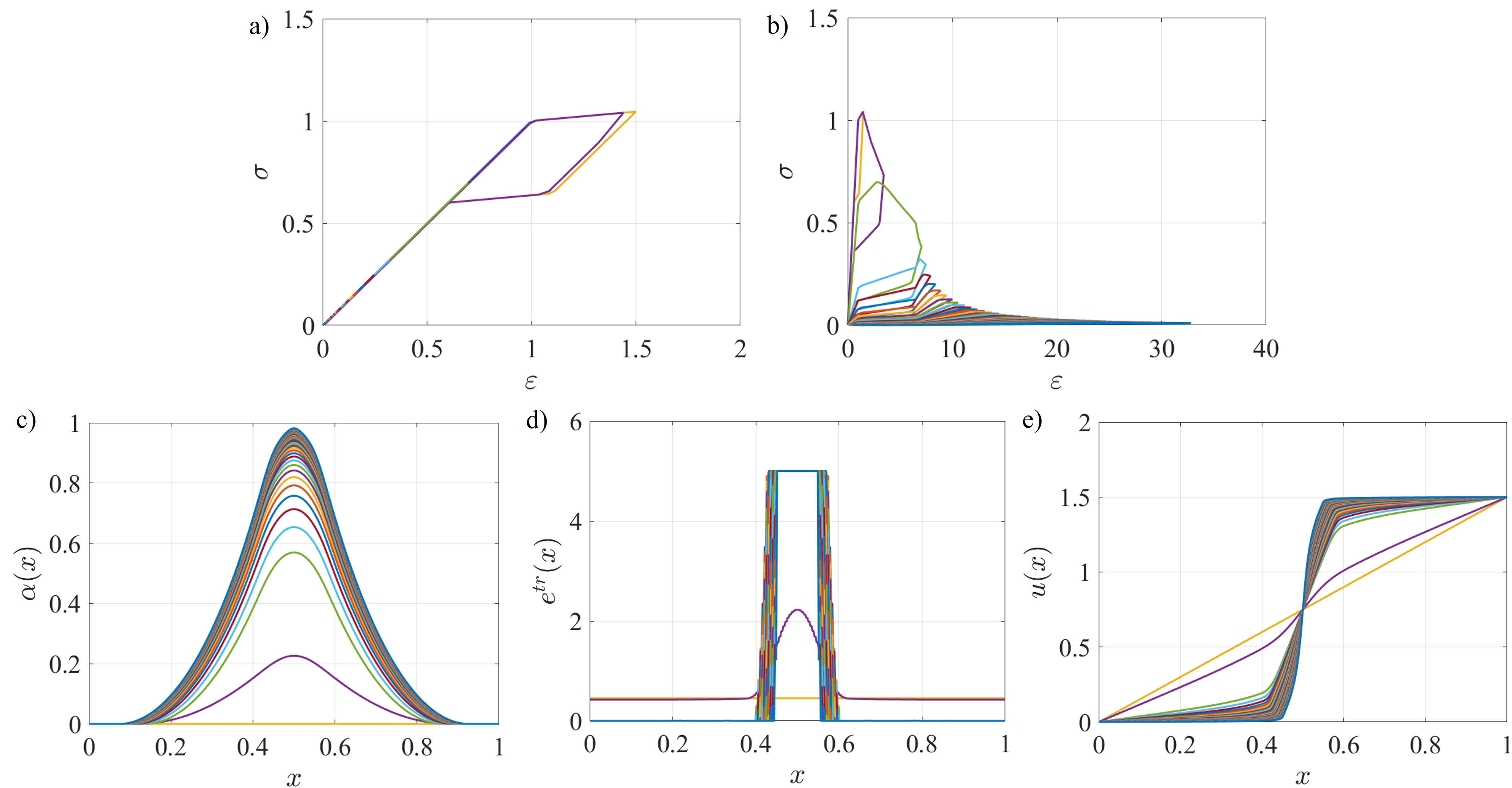}
    \caption{Localized response for the E-T-TD model with $\varepsilon_L =5$ and U ranging between 0 and 1.5: a) local stress-strain curves at the last element under fatigue cycles; b) local stress-strain curves at the central element under fatigue cycles; c) damage profile along the bar at the peak stress of subsequent fatigue cycles; d) transformation strain profile along the bar at the peak stress of subsequent fatigue cycles; e) displacement profile along the bar at the peak stress of subsequent fatigue cycles.}
    \label{fig12}
\end{figure*}

In the second example, we adopt a finite value $\varepsilon_L =5$ to model the saturation of the transformation strain in the fully martensitic phase, keeping the boundary conditions and all other parameters fixed. The results are reported in Figure \ref{fig12}. The material response initially follows the behavior described in the first example, until the transformation strain at the center of the bar reaches $\varepsilon_L$. At this stage, the material enters the martensitic phase, and the subsequent damage evolution proceeds along the elastic martensitic loading path. In general, the morphology of the local stress-strain cycles may vary depending on the interplay between damage and transformation strain, as depicted in Figure \ref{fig12}b which shows the response of the central element of the bar during successive loading cycles. Additional stages can be observed, corresponding to elastic loading and unloading within the fully martensitic phase. 
Since the transformation strain cannot locally increase beyond $\varepsilon_L$, the minimization of the total energy leads to a slight broadening of the damage profile, resulting in a smoother gradient near the centerpoint of the bar (Figure \ref{fig12}c). The strain is uniformly equal to $\varepsilon_L$ around the center to accommodate the applied deformation, and it progressively decreases to zero at the boundaries of the bar due to pseudoelastic recovery. Accordingly, the displacement jump is slightly diffused in the localization area and the displacement profile progressively flattens at the ends of the bar (Figure \ref{fig12}e). Assuming a threshold of 0.01 on the peak stress, the fatigue life slightly increases compared to the case with $\varepsilon_L \to \infty$ (43 cycles versus 27 cycles).

\begin{figure*}[thb!]
    \centering
    \includegraphics[width=\textwidth]{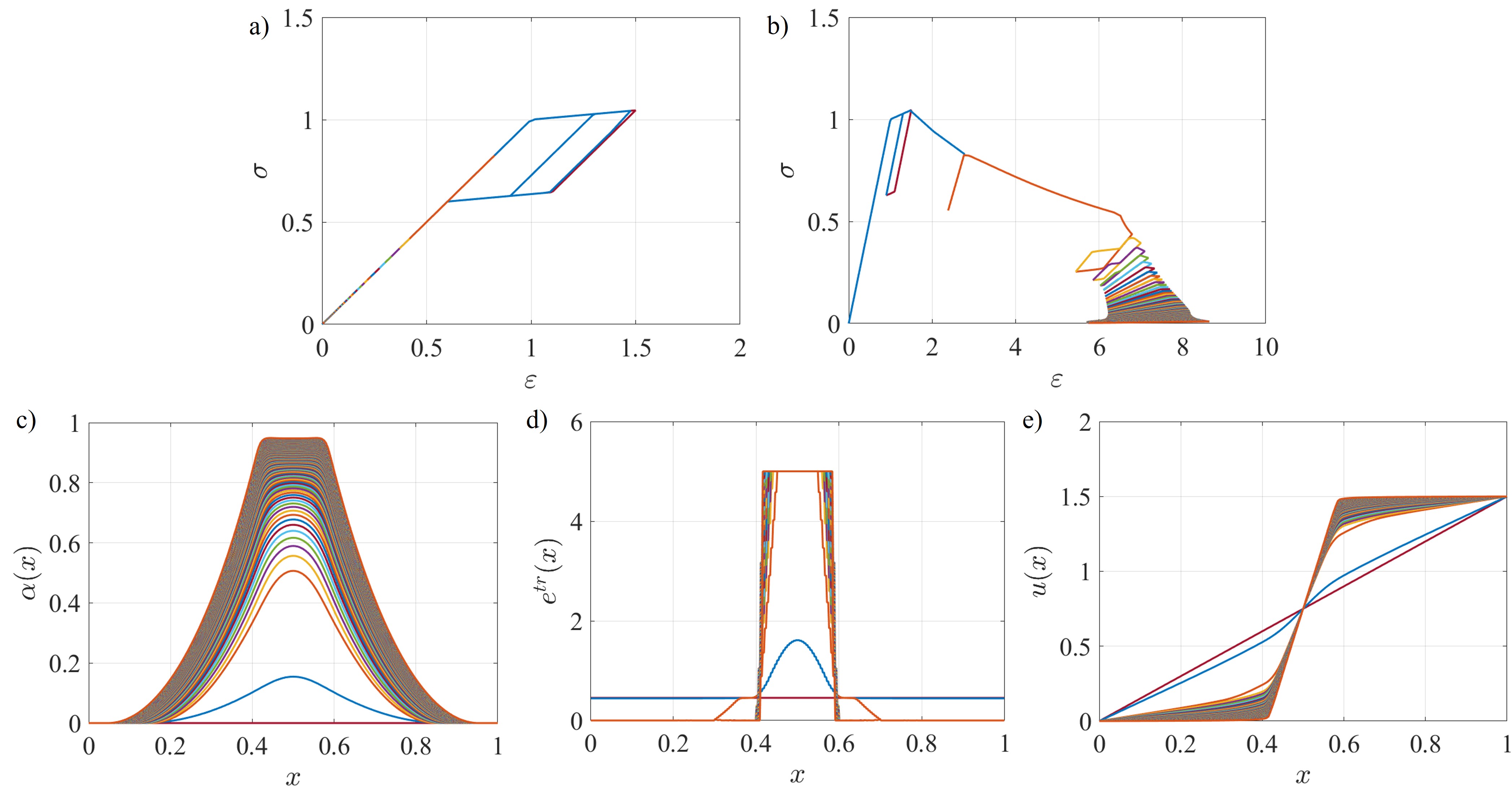}
    \caption{Localized response for the E-T-TD model with $\varepsilon_L =5$ and U ranging between 0.9 and 1.5: a) local stress-strain curves at the last element under fatigue cycles; b) local stress-strain curves at the central element under fatigue cycles; c) damage profile along the bar at the peak stress of subsequent fatigue cycles; d) transformation strain profile along the bar at the peak stress of subsequent fatigue cycles; e) displacement profile along the bar at the peak stress of subsequent fatigue cycles.}
    \label{fig13}
\end{figure*}

The effects of the saturation of the transformation strain are further emphasized considering partial displacement cycles between a minimum value larger than 0 and $U_0$, representing the most common case in the cardiovascular field. For example, we apply a monotonic increasing displacement up to 1.5, followed by unloading to 0.9 and then loading-unloading half-cycles between 0.9 and 1.5, obtaining the results in Figure \ref{fig13}. 
Similar considerations to the previous examples can be made regarding the local stress-strain cycles. However, in this case, the transformation strain is not fully recovered during unloading, leading to a residual deformation energy given by elastic and inelastic contributions. Under cyclic loads, the system progressively evolves toward a configuration in which the maximum damage is diffused over a wider region around the centerpoint of the bar, which is more convenient from an energetic point of view. The resulting damage profile (Figure \ref{fig13}c) is flattened on the support region and decreases to zero at the boundaries. The width of the support region progressively increases under cyclic loads and stabilizes for high damage values. The transformation strain is uniformly equal to $\varepsilon_L$ in the support region and the displacement shows a linear variation (Figures \ref{fig13}d and \ref{fig13}e). Assuming a threshold on the peak stress of 0.01, we compute a fatigue life of 184 cycles. Neglecting the saturation of the transformation strain ($\varepsilon_L \to \infty$), the damage would assume the standard profile with the maximum value localized at the singular point and the transformation strain locally tending to infinity. This unrealistic hypothesis would result in a lower fatigue life, equal to 64 cycles in our example. Therefore, the limit strain $\varepsilon_L$ plays a major role in distributing damage over a wide region where the material is fully transformed. As a consequence, a larger portion of the bar contributes to resisting fracture, delaying the damage process.
\section{Case study: uniaxial fatigue of Ni-Ti multi-wire samples} \label{sec6}
In this section, our proposed model is applied to a real case study for analyzing the response of Ni-Ti multi-wire samples undergoing uniaxial fatigue loading. The samples, depicted in Figure \ref{fig14}a, present a tubular geometry, with nine wires along the circumference. These were obtained from tubes used to produce cardiovascular devices by laser cutting, followed by chemical etching and a proper sequence of thermal treatments \citep{Brambilla2024}. The design of the samples is optimized such that the strain is localized in the wires, avoiding any singularities in the transition region near the fillets. Therefore, each wire can be modeled as a 1D straight bar undergoing a uniaxial stress-strain state, and the proposed phase transformation-damage model can be applied. The average final dimensions of the samples are sketched in Figure \ref{fig14}a, with a wire width of 0.38 mm, a thickness of 0.2 mm, and a gage length of 14.47 mm. We calibrate the model parameters from available experimental data and study the fatigue response of the wires. To compute the fatigue life, we adopt different criteria based either on the minimum stress or on the maximum damage. For several loading conditions, we compare the fatigue life obtained from the model with the experimental outcomes.
\subsection{Experimental data}
Both static tensile data and uniaxial fatigue data are available from previous experimental campaigns on multi-wire samples \citep{Brambilla2024}. All tests were carried out by testing the samples in water at a constant temperature of 37 °C, at which the material shows a characteristic pseudoelastic behavior, exploiting an ad hoc setup described in \cite{Brambilla2024}. 

\begin{figure*}[t!]
    \centering
    \includegraphics[width=0.7\textwidth]{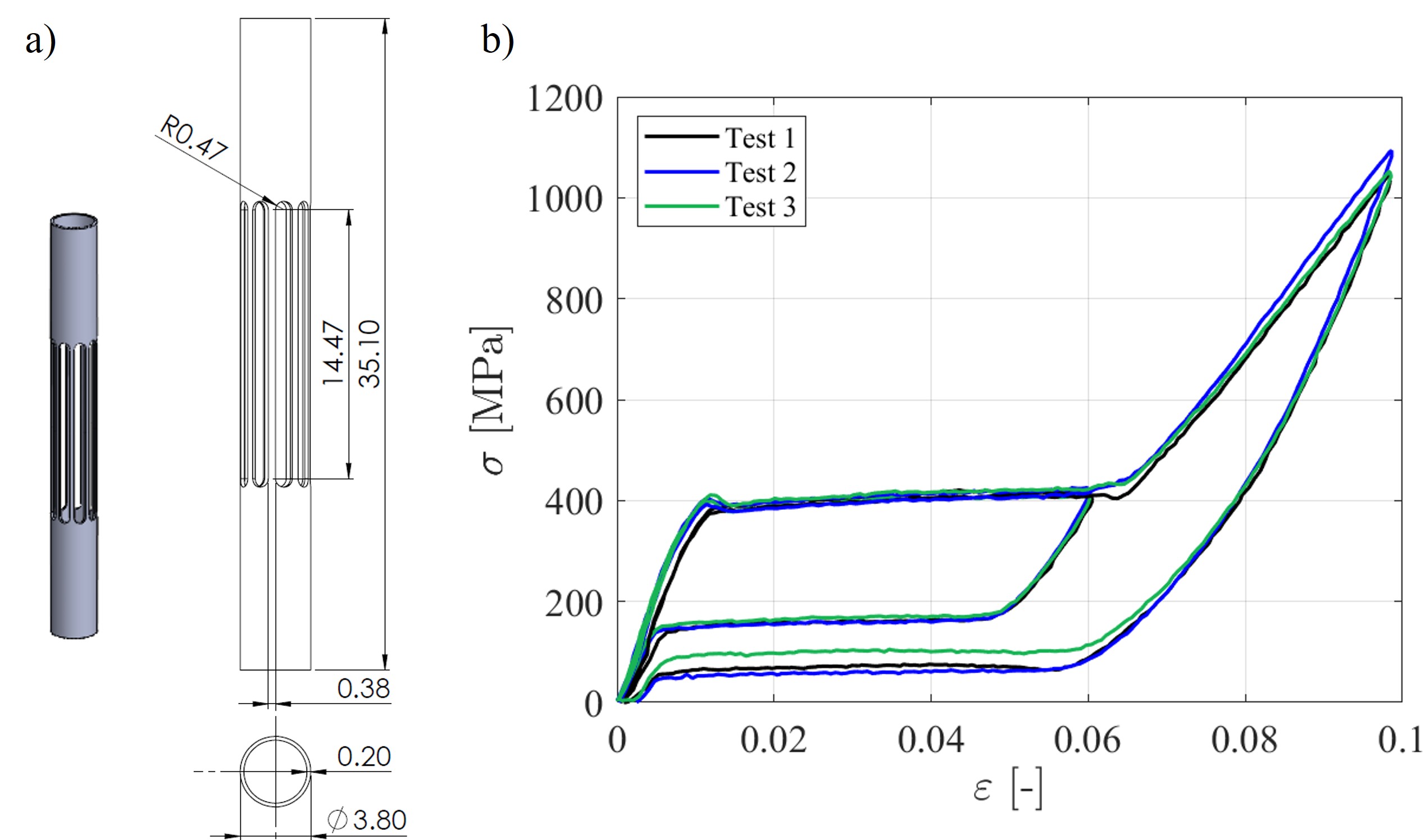}
    \caption{a) Three-dimensional representation and technical drawing of the multi-wire samples analyzed in this study. All dimensions are expressed in millimeters. b) Experimental true stress-strain curve of three multi-wire samples.}
    \label{fig14}
\end{figure*}

Three samples were tested in tension in quasi-static conditions under axial displacement control. Each sample was loaded up to a nominal strain of $6\%$, followed by unloading and by a second loading-unloading cycle up to $10\%$. The stress-strain curve reported in Figure \ref{fig14}b was obtained from the machine load-displacement data knowing the sample cross-section and gage length.

Regarding the fatigue campaign, multi-wire samples were tested in axial displacement control at several combinations of mean strain and strain amplitude. The strain was computed based on the applied displacement, assuming homogeneous strain. Each sample was preloaded up to a strain of $6\%$, followed by unloading at the mean strain level of the fatigue cycle. Then, cyclic loads at the desired strain amplitude were applied with a frequency of 20 Hz, assuming a run-out of 1 million cycles. This loading sequence is commonly adopted for biomedical applications of SMAs to mimic the loading process of cardiovascular devices. Among the tests reported in \cite{Brambilla2024}, only conditions up to a mean strain of $5.5\%$ were considered, which are the most relevant considering real applications. Table \ref{tab6} reports for each test the mean strain, the strain amplitude, and the experimental range of the number of cycles to failure, considering the nine wires of each sample. Regarding test 6, only two fractures were observed at the sample fillets probably due to manufacturing defects, while the other seven wires reached the run-out. Therefore, test 6 can be reasonably classified as a run-out condition. In all other cases, fractures were observed along the wires.
\begin{table*} [t]
\centering
\caption{Experimental fatigue campaign carried out on multi-wire samples in terms of mean strain $\varepsilon_m$, strain amplitude $\varepsilon_a$, and experimental range of number of cycles to failure $N_f$. Comparison between the experimental and the predicted fatigue lives.}
 \begin{tabular}{|wc{2em}|  wc{3em}| wc{3em}|  wc{4em} wc{4em}| wc{6.5em} wc{6.5em} wc{6.5em}|}
\hline 
& & & \multicolumn{2}{c|}{\textbf{Experimental result}} & \multicolumn{3}{c|}{\textbf{Numerical prediction}}\\  
\multirow{-2}{*}{\textbf{Test}} & \multirow{-2}{*}{$\varepsilon_m$ [\%]} & \multirow{-2}{*}{$\varepsilon_a$ [\%]} & $N_{f_{min}}$ & $N_{f_{max}}$ & $\sigma_{th}=0.05$ MPa & $\sigma_{th}=0.01$ MPa & $\alpha_{th}=0.99$ \\
\hline
1 & 1.5 & 0.3 & \multicolumn{2}{c|}{Run-out} &   \multicolumn{3}{c|}{Run-out} \\
2 & 1.5 & 0.5 & 4300 & 6800 & 5440 & 12221 & 9820\\
3 & 1.5 & 1 & 2400 & 3200 & 1345 & 2980 & 2500 \\
4 & 2 & 0.7 & 3900 & 6100 & 3628 & 8138 & 6599\\
5 & 2.5 & 0.3 & \multicolumn{2}{c|}{Run-out} &  \multicolumn{3}{c|}{Run-out}\\
6 & 2.5 & 0.4 & 18000 & Run-out &  \multicolumn{3}{c|}{Run-out}\\
7 & 2.5 & 0.6 & 4300 & 14100 & 8055 & 17985 & 14463 \\
8 & 2.5 & 2 & 1700 & 2800 & 1003 & 2240 & 1909 \\
9 & 3 & 2 & 2100 & 2700 & 1190 & 2664 & 2248\\
10 & 4 & 0.5 & \multicolumn{2}{c|}{Run-out} &   \multicolumn{3}{c|}{Run-out}\\
11 & 4 & 0.7 & 3200 & 12400 & 12961 & 29066  & 23256\\
12 & 4 & 1 & 4700 & 5700 & 4707 & 10563 & 8536\\
13 & 5.5 & 0.5 & \multicolumn{2}{c|}{Run-out} &   \multicolumn{3}{c|}{Run-out}\\
\hline
\end{tabular}
\label{tab6}
\end{table*}
\subsection{Model calibration}
\begin{figure*}[thb!]
    \centering
    \includegraphics[width=0.8\textwidth]{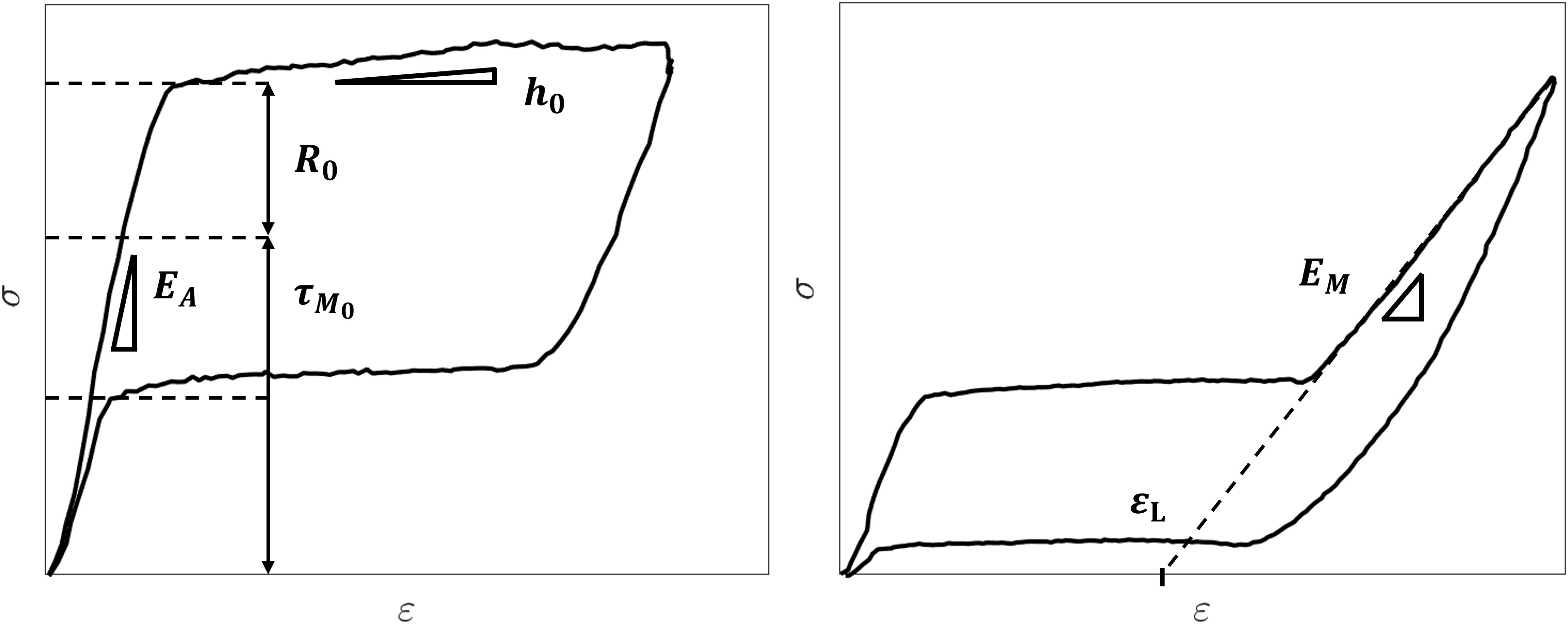}
    \caption{Calibration of the model parameters from the experimental stress-strain curves.}
    \label{fig15}
\end{figure*}

For each static stress-strain curve, the model parameters needed to describe the material response with no damage are calibrated as graphically explained in Figure \ref{fig15}. The elastic modulus of the austenitic phase ($E_A$) is obtained as the slope of the initial elastic loading stage, whereas the elastic modulus of martensite ($E_M$) is computed as the slope of the elastic loading path in the fully martensitic phase. The elastic modulus of the material, modeled as a composite of the austenitic and martensitic phases, is defined as a function of the transformation strain according to the formulation proposed by \cite{Auricchio2009}:
\begin{equation} \label{eq51}
    E(e^{tr}) = \frac {\varepsilon_L}{\frac{\varepsilon_L - \lvert e^{tr} \rvert}{E_A} + \frac{\lvert e^{tr} \rvert}{E_M}}.
\end{equation}
In this way, when $e^{tr}=0$, the elastic modulus corresponds to the one of the austenitic phase, whereas, when $e^{tr}=\varepsilon_L$, it is equal to the one of the martensitic phase. This formulation resembles the Reuss scheme adopted for composite materials, assuming platelet inclusions in a periodic composite, which represents the most feasible choice for SMAs from an experimental point of view \citep{Auricchio1997}. We acknowledge that introducing a dependence of the elastic modulus on the transformation strain $e^{tr}$, as defined in (\ref{eq51}), compromises the variational consistency of our proposed model in its current form. However, this dependency is necessary to accurately describe the cyclic response of the material under specific combinations of mean strain and strain amplitude. Different loading conditions can result in elastic cycles in fully austenitic or martensitic states, as well as in mixed-phase configurations, or in hysteretic cycles involving partial phase transformations. Assuming a fixed elastic modulus regardless of phase composition would prevent an accurate representation of elastic cycles, thereby leading to unreliable numerical predictions. Despite the loss of variational consistency, we will show that the resulting material behavior is consistent with the results obtained using the variational model in Section \ref{sec5-2}.

The material parameter $\tau_{M_0}$ is calculated as the mean between the stress at the beginning of the forward transformation and the stress at the end of the reverse transformation, while $R_0$ is half the difference between the two stresses. The slope of the forward transformation path is used to compute the hardening modulus $h_0$, and $\varepsilon_L$ is obtained as the maximum transformation strain in the fully martensitic phase. All material parameters are set to the average of the values obtained from the three static tensile tests. 

Concerning the fracture properties, the fracture toughness $G_c$ is taken from the literature as the value $J_{I_c}$ measured on Ni-Ti CT samples according to an ad hoc procedure for SMAs \citep{Haghgouyan2019}, equal to 120 kJ/m$^2$. The specific fracture energy $w_1$ is calibrated numerically, taking advantage of the relationship between $G_c$ and $w_1$ through the internal length $l$ (\ref{eq46}). For this purpose, the localized response of a bar with length and cross-section equal to those of a single wire of the multi-wire samples is simulated for various internal lengths $l$, corresponding to different specific energies $w_1$. A monotonically increasing displacement is applied at the end of the bar up to an average strain of $20\%$. The deformation reached at fracture is compared with the experimental results reported in the literature for different sample geometries and manufacturing conditions \citep{Rainer2009, Mwangi2020, Marandi2021}. An internal length of 0.12 mm is selected, corresponding to a specific fracture energy of 375 MPa and a deformation at fracture of approximately $17\%$. 
Table \ref{tab7} summarizes the model parameters resulting from the calibration procedure, which are adopted in the following for the numerical simulations of fatigue.
\begin{table*}[th!]
\centering
\caption{Parameters of the proposed model calibrated from experimental tests on multi-wire samples and numerical simulations of quasi-static tension.}
\begin{tabular}{c c c c c c c c} 
\hline
$E_A$ [MPa] & $E_M$ [MPa]&  $\tau_{M_0}$ [MPa] & $R_0$ [MPa] & $h_0$ [MPa] & $\varepsilon_L$ [\%]& $w_1$ [MPa] & $l$ [mm]\\
\hline 
45000 & 20000 & 255 & 125 & 605 & 4.52 & 375 & 0.12 \\ 
\hline
\end{tabular}
\label{tab7}
\end{table*}
\subsection{Numerical results}
We study a straight bar with length $L$ and cross-section $A$ equal to those of a single wire of the multi-wire samples. A cyclic displacement is applied at $x=L$, mimicking the loading procedure of the uniaxial fatigue tests. We simulate all the experimental tests in Table \ref{tab6}, converting the boundary conditions in terms of nominal strain into the correspondent displacements, exploiting the relation $\overline{\varepsilon} = U/L$. 
The problem is solved numerically as described in Section \ref{sec5-2}. The bar is discretized in 500 elements with uniform length, and the time interval of each half-cycle is divided in 50 time steps. The model parameters in Table \ref{tab7} are adopted in all simulations. We account for the saturation of the maximum transformation strain at $\varepsilon_L$ and we consider specific elastic properties for any mixture of austenitic and martensitic phases. For the softening parameter, we adopt $s=2$, selected based on a sensitivity analysis. Higher values, such as $s=3$, were also investigated in a preliminary study but consistently led to non-conservative fatigue life predictions and were therefore excluded from further consideration.
The fatigue life for each condition is computed by fixing a threshold either on the minimum stress, $\sigma_{th}$, or on the maximum damage, $\alpha_{th}$. A sensitivity analysis is carried out considering thresholds of 0.05 MPa and 0.01 MPa for the stress, and 0.99 for the damage.

Under the loading conditions where the experiments lead to run-out (10$^6$ cycles with no failure), the model correctly predicts an infinite fatigue life. In these cases, after the first preloading cycle, the bar undergoes elastic loading-unloading cycles in a mixture of austenitic and martensitic phases. The phase content of the material remains constant since no phase transformation occurs during cyclic loads. Therefore, the transformation strain and the accumulated transformation strain remain constant as well, equal to the value reached at the end of the preload. As a consequence, no damage evolution occurs and $\alpha$ remains uniformly equal to zero everywhere, resulting in an infinite fatigue life. 
These considerations apply to tests 1, 5, 6, 10, and 13, in agreement with experimental observations.
Note that, in the past years, the fatigue assessment of Ni-Ti cardiovascular devices was mainly based on the strain limit curves provided by the seminal work by \cite{Pelton2008}, without addressing the local deformation regime of the material. Only recently, some studies \citep{Brambilla2024, Launey2023, Catoor2019} recognized the role of the deformation regime on the fatigue behavior, distinguishing between nominally elastic cycles, generally associated to an infinite fatigue life, and hysteretic cycles with phase transformation.
Accordingly, some ad hoc fatigue index parameters based on the dissipated energy \citep{Launey2023} or the cyclic stress amplitude normalized with respect to the amplitude between the stress plateaus \citep{Catoor2019} have been proposed to properly discriminate between safe and failure conditions. Interestingly, the proposed gradient damage model allows to correctly classify elastic fatigue cycles as safe thanks to its inherent formulation, reproducing the key feature of recent ad hoc fatigue criteria for SMAs.

\begin{figure*}[thpb!]
\centering
\includegraphics[width=\textwidth]{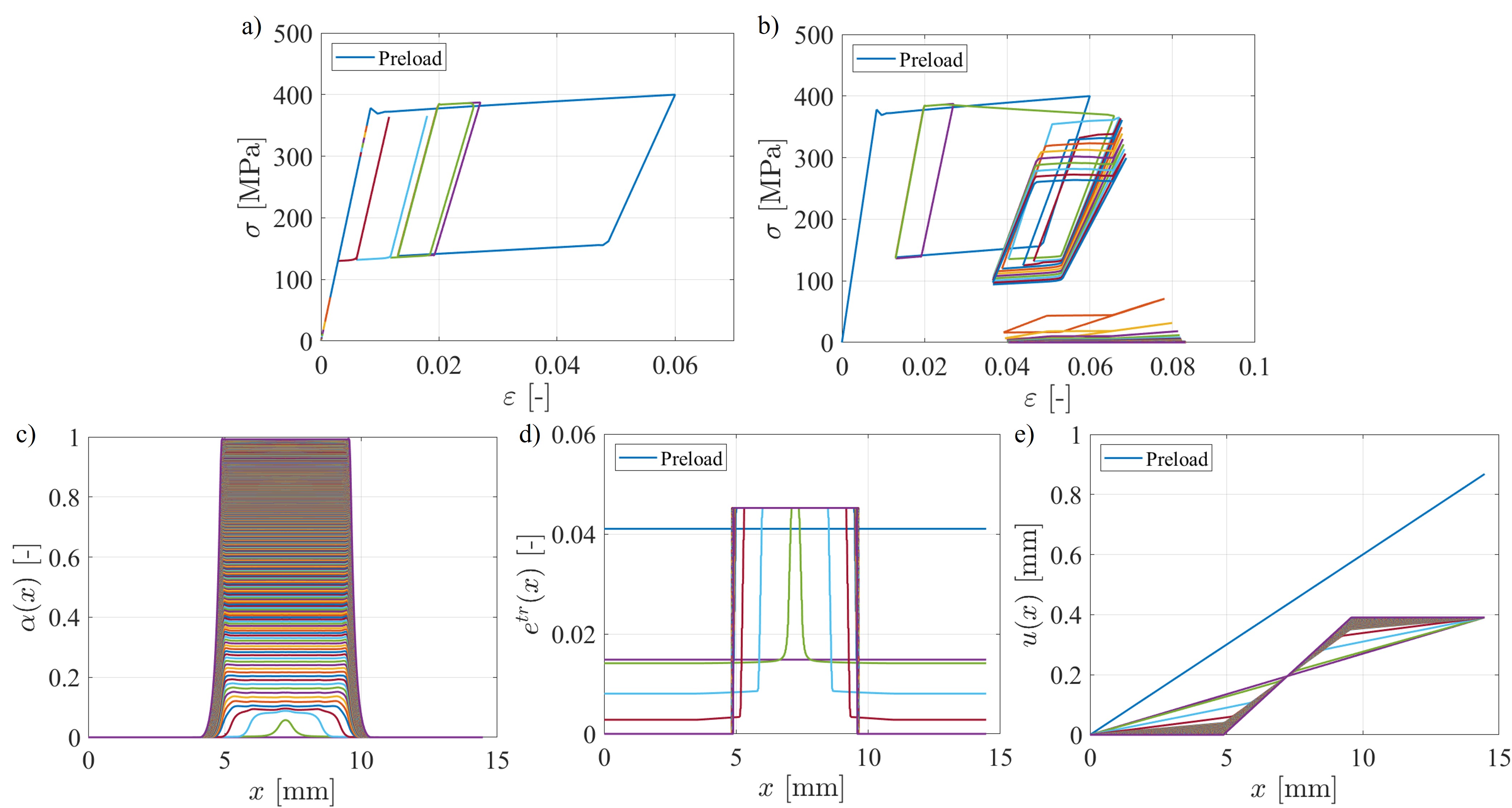}
\caption{Localized response for test 4: a) local stress-strain curves at the last element; b) local stress-strain curves at the central element; c) damage profile along the bar at the peak stress of subsequent fatigue cycles; d) transformation strain profile along the bar at the peak stress of subsequent fatigue cycles; e) displacement profile along the bar at the peak stress of subsequent fatigue cycles. For the sake of clarity, stress-strain curves are shown for all cycles up to 10 cycles after damage initiation, and one cycle every 100 cycles thereafter.}
\label{fig16}
\end{figure*}

\begin{figure*}[thpb!]
    \centering
    \includegraphics[width=\textwidth]{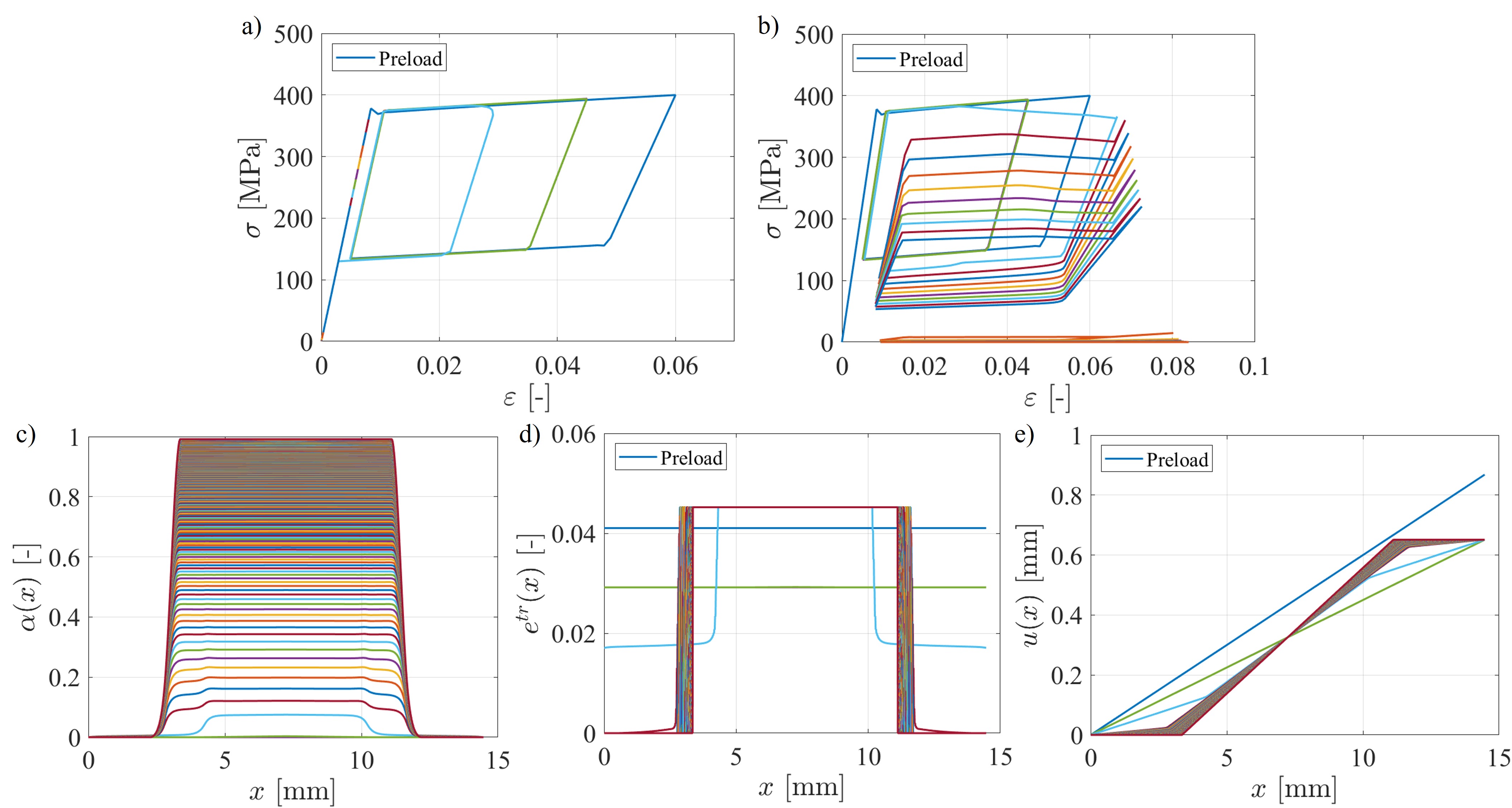}
\caption{Localized response for test 8: a) local stress-strain curves at the last element; b) local stress-strain curves at the central element; c) damage profile along the bar at the peak stress of subsequent fatigue cycles; d) transformation strain profile along the bar at the peak stress of subsequent fatigue cycles; e) displacement profile along the bar at the peak stress of subsequent fatigue cycles. For the sake of clarity, stress-strain curves are shown for all cycles up to 10 cycles after damage initiation, and one cycle every 100 cycles thereafter.}
    \label{fig17}
\end{figure*}

Let us now consider the loading conditions where a finite fatigue life is obtained; tests 4 and 8 can be considered representative for low and high strain amplitude conditions, see Figures \ref{fig16} and \ref{fig17}. For each case, we report the local stress-strain response in the elements at the end and at the center of the bar, together with the damage, transformation strain, and displacement profiles. 
The same considerations of the examples in Section \ref{sec5-2} can be made for the local stress-strain response. In all cases, the regions at the ends of the bar are rapidly unloaded, showing just a few hysteretic cycles followed by elastic cycles in austenite (see for example Figure \ref{fig16}a). Conversely, at the center of the bar, the material displays hysteretic stress-strain cycles. The mechanical hysteresis increases with the nominal strain amplitude, as depicted in Figures \ref{fig16}b and \ref{fig17}b where stress-strain cycles are shown cycle-by-cycle up to 10 cycles after damage initiation, and one cycle every 100 cycles thereafter. Damage initiation occurs at different numbers of cycles according to the severity of the loading conditions (124 cycles for test 4 and 26 cycles for test 8). The damage initially assumes the standard profile with a peak value at the centerpoint of the bar. The combination of boundary conditions and material parameters is such that, in all cases, as soon as the transformation strain reaches $\varepsilon_L$, the system assumes the most energetically convenient configuration spreading the maximum damage on a wide support region around the centerpoint. Therefore, similar considerations to Section \ref{sec5-2} can be made for the damage, transformation strain and displacement profiles. As $e^{tr}$ saturates, elastic loading-unloading paths in the fully martensitic phase can be observed in the local stress-strain cycles at the center of the bar. The width of the support region progressively increases under cyclic loads, with more and more portions of the bars in fully transformed conditions, and it stabilizes for high damage levels. Moreover, the final width increases with the nominal strain, reaching e.g. approximately 5 mm for test 4 ($\varepsilon_m=2\%$, $\varepsilon_a=0.7\%$) and 10 mm for test 8 ($\varepsilon_m=2.5\%$, $\varepsilon_a=2\%$). 

The existence of a limit value for the transformation strain is a crucial feature that distinguishes the fatigue behavior of SMAs from that of conventional elastoplastic materials. Some preliminary simulations under the unrealistic hypothesis $\varepsilon_L \to \infty$ predicted failure in a few hundreds of cycles, strongly underestimating the experimental fatigue lives. The saturation of the transformation strain, by diffusing damage over a wide region where the material is fully transformed, results in higher fatigue lives which are more realistic compared to experimental outcomes. 

The predicted numbers of cycles to failure are reported in Table \ref{tab6} for different choices of the threshold, based on either a minimum stress or a maximum damage criterion. Unfortunately, the results reveal a strong sensitivity of the predictions to the selected threshold, and it is not possible to identify a single value that is applicable across all tests. For conditions characterized by low mean strain and high strain amplitude (tests 3, 8, and 9), a stress threshold $\sigma_{th}=0.01$ MPa yields predictions that best match the experimental range of cycles to failure. In contrast, higher thresholds lead to more conservative estimates. 
For the other conditions, a stress threshold $\sigma_{th}=0.05$ MPa provides the closest agreement with the average experimental fatigue life. 
Using a damage threshold $\alpha_{th}=0.99$ results in numerical predictions that fall between those obtained with $\sigma_{th} = 0.01$ MPa and $\sigma_{th} = 0.05$ MPa, representing a balanced compromise across most test conditions. 
Overall, regardless of the selected threshold, the mismatch between the predicted and the average experimental fatigue life is within a factor of three. Additionally, the model correctly captures the increasing trend in the fatigue resistance at fixed strain amplitude with increasing mean strain reported in the literature \citep{Robertson2012}. However, the available experimental data display a binary behavior, with specimens either failing within 20000 cycles or reaching the run-out. At this stage, this limitation prevents a clear assessment of whether the method can accurately predict the order of magnitude of fatigue life across the full range from 10$^3$ to 10$^6$ cycles.

Clearly, the comparison between experimental and numerical fatigue lives may be affected by experimental uncertainties. First, strain boundary conditions for each test are derived from machine data due to the infeasibility of local measurement methods. This might lead to a mismatch between the simulated response and the actual material behavior in the wire gage length. Moreover, the material response varies from test to test due to the intrinsic variability in the material and the differences arising from thermomechanical processing. These factors can lead to deviations from the average properties assumed in the numerical model, contributing to potential mismatches in fatigue life predictions. 
Taking these aspects into account, a more extensive experimental campaign would be required to better evaluate the predictive capability of the model, including intermediate loading conditions and minimizing variability in both material properties and testing conditions. Furthermore, based on the current results, the constitutive equations governing stress softening with damage warrant further investigation, potentially exploring formulations specifically tailored for SMAs.
\section{Conclusions}
This study proposes a novel variational phase-field model for fracture and fatigue in SMAs. The model is developed in a 1D setting and captures the key features of the pseudoelastic behavior at a fixed temperature.
A coupling between phase transformation and damage is introduced through exploratory constitutive functions for the material parameters of the SMA, resulting in a softening response.
We analyze the homogeneous and localization responses of a 1D bar under monotonically increasing end displacement, considering various parameter sets that correspond to different macroscopic behaviors. 
The inherent structure of the model, linking damage evolution with the accumulated transformation strain, enables to describe fatigue effects under cyclic loading without the need for additional variables. 
A key feature of the model is the introduction of a limit value for the transformation strain, beyond which the material is fully martensitic and exhibits an elastic behavior. This saturation leads to a distinctive response: the localized damage is diffused over a broader region where the material is fully transformed, thereby delaying the fracture process compared to standard elastoplastic materials.

The model is applied to simulate the uniaxial fatigue response of Ni-Ti multi-wire samples used in the cardiovascular field. Notably, the model effectively distinguishes between safe and critical loading conditions, capturing the observed trend of increased fatigue resistance with higher mean strain at a fixed strain amplitude. However, it does not accurately predict the number of cycles to failure across all conditions, revealing a strong sensitivity of the results to the selected threshold variable.

Although currently limited to a simplified 1D setting, this work represents the first systematic effort to compare a phase-field model for Ni-Ti SMAs with experimental fatigue data, highlighting both the potential and current limitations of the approach. The results suggest that further refinement of the modeling choices, as well as an increase of the experimental dataset, is still required to fully assess the predictive capability of the proposed model and to evaluate its extension to more complex three-dimensional scenarios for the fatigue assessment of SMA components in practical engineering applications.
\backmatter
\bmhead{Acknowledgements}
This study was supported in part by IDEA League through a Short-term Research Exchange Grant. 
LDL would like to acknowledge funding from the Swiss National Science Foundation through Grant No. 200021-219407 ‘Phase-field modeling of fracture and fatigue: from rigorous theory to fast predictive simulations’.

\appendix
\numberwithin{equation}{section}
\section{Variational formulation of SMA constitutive equations}
\label{appendix:A}
Assuming that the variational formulation holds, let us prove that (\ref{eq8})-(\ref{eq9}) imply (\ref{eq11})-(\ref{eq14}). For a given state $(\varepsilon, e^{tr}, \overline{e^{tr}})$, consider a test direction $e^{tr^*}=e^{tr} + a \delta e $, with $0 < a \leqslant 1$ and $\delta e  \in \mathbb{R}$. Expanding the total energy density around the state $(\varepsilon, e^{tr}, \overline{e^{tr}})$ up to first order in $a$ yields
\begin{equation} \label{eq52}
\begin{split}
&W_T(\varepsilon, e^{tr}+ a \delta e,  \overline{e^{tr}} + a \lvert \delta e \rvert) = W_T(\varepsilon, e^{tr},  \overline{e^{tr}}) \\&- E_0 (\varepsilon - e^{tr}) a \delta e + (\tau_{M_0} + h_0 \lvert e^{tr}\rvert) \frac{\partial \lvert e^{tr}  \rvert}{\partial e^{tr}} a \delta e \\&+ R_0 a \lvert \delta e \rvert.
\end{split}
\end{equation}
Invoking the stability criterion (\ref{eq8}) and dividing by $a$ gives
\begin{equation}  \label{eq53}
- \sigma \delta e + (\tau_{M_0} + h_0 \lvert e^{tr}\rvert) \frac{\partial \lvert e^{tr}  \rvert}{\partial e^{tr}} \delta e + R_0 \lvert \delta e \rvert \geq 0.
\end{equation}
Dividing by $\lvert \delta e \rvert$ results in 
\begin{equation} \label{eq54}
\frac{\delta e}  {\lvert \delta e \rvert} \biggl[ \sigma - (\tau_{M_0} + h_0 \lvert e^{tr}\rvert) \frac{\partial \lvert e^{tr}  \rvert}{\partial e^{tr}} \biggr]  - R_0 \leq 0,
\end{equation} 
or 
\begin{equation} \label{eq54_bis}
 \text{sign} (\delta e) X \leq R_0.
\end{equation} 
Considering the energy balance condition and developing (\ref{eq9}) through the differentiation chain rule, we obtain
\begin{equation} \label{eq55}
- \sigma {\dot{e}^{tr}} + (\tau_{M_0} + h_0 \lvert e^{tr}\rvert) \frac{\partial \lvert e^{tr}  \rvert}{\partial e^{tr}} {\dot{e}^{tr}} + R_0 \lvert  {\dot{e}^{tr}} \rvert = 0.
\end{equation} 
Introducing the transformation stress $X$ and dividing by $\lvert   {\dot{e}^{tr}} \rvert$ gives
\begin{equation} \label{eq56}
X \frac{ {\dot{e}^{tr}}} {\lvert {\dot{e}^{tr}} \rvert}  - R_0 = 0.
\end{equation}
From (\ref{eq54_bis}) and (\ref{eq56}) we obtain 
\begin{equation} \label{eq57}
{\dot{e}^{tr}} \left\{ \begin{array}{rcl} \geq 0 & \mbox{if} & X = R_0 \\ =0 & \mbox{if} & \lvert X \rvert -R_0 < 0 \\ \leq 0 & \mbox{if} & X = -R_0 \end{array}\right.,
\end{equation}
which are equivalent to (\ref{eq11})-(\ref{eq14}).
\section{Construction of the evolution problem from the energetic principles}
\label{appendix:B}
\subsection{The stability condition}
Dividing (\ref{eq31}) by $a$ and passing to the limit when $a \to 0$ provides the first order stability conditions:
\begin{equation}  \label{eq58}
\begin{split}
&\frac {d}{da}\mathcal{W} (u_t + av, e^{tr}_t + aq,  \overline{e^{tr}}_t + a\lvert q \rvert , \alpha_t + a\beta) \bigg\rvert _{a=0} \\& \geq 0 \ \ \  \forall (v, q, \beta) \ \text{admissible}.
\end{split}
\end{equation}
Considering the virtual state $\xi^* =(u^*, e^{tr^*}, \overline{e^{tr}}^*, \alpha^*)  = \xi_t + a(v, q, \lvert q \rvert, \beta)$, characterized by the singular set $J = J(u_t) \cup J(v) $, the virtual strain and the virtual transformation strain fields are given by
\begin{equation} \label{eq59}
\varepsilon^* = \varepsilon_t + av'(x) + a \sum_{x_i \in J(v)}[\![ v]\!](x_i) \delta_{x_i}, 
\end{equation}
\begin{equation} \label{eq60}
e^{tr^*} = e^{tr}_t + aq^R(x)  + a \sum_{x_i \in J(v)}[\![ v]\!](x_i) \delta_{x_i}. 
\end{equation}
\begin{equation} \label{eq60bis}
\overline{e^{tr}}^* = \overline{e^{tr}}_t + a \lvert q^R(x) \rvert  + a \sum_{x_i \in J(v)}\lvert {[\![ v ]\!]}(x_i)\rvert \delta_{x_i}. 
\end{equation}

Using the definition of energy (\ref{eq29}) gives
\begin{equation} \label{eq61}
\begin{split}
&\int_{[0,L]\backslash (J(u)\cup J(v))}  \biggl(  \sigma (v'-q^R) \\&+ (\tau_M(\alpha) + h(\alpha) \lvert e^{tr^R} \rvert)  \ \text{sign} (e^{tr^R}) q^R + \\& R(\alpha) \lvert q^R \rvert \biggr) dx \\
&+ \int_{[0,L]\backslash J(u)}  \biggl[   \biggl( -\frac{1}{2} S'(\alpha) \sigma^2 + \tau_M'(\alpha) \lvert e^{tr^R} \rvert \\& + \frac{1}{2} h'(\alpha) \lvert e^{tr^R} \rvert^2 + R'(\alpha) \overline{e^{tr^R}} + w'(\alpha) \biggr) \beta  \\& +2w_1 l^2 \alpha' \beta'  \biggr] dx \\
&+ \sum_{J(u)} \biggl( (\tau_M(\alpha) + h(\alpha)  \lvert [\![ u]\!] \rvert ) \ \text{sign} ([\![ u]\!])q^R \\& + R(\alpha) \lvert q^R \rvert  \biggr)\\
&+ \sum_{J(u)} \biggl[ \biggl( \tau_M'(\alpha) \lvert [\![ u]\!] \rvert + \frac{1}{2} h'(\alpha) \lvert [\![ u]\!] \rvert^2 + R'(\alpha) \overline {[\![ u ]\!]} \biggr) \beta \biggr] \\
& + \sum_{J(v)} \biggl( ( \tau_M(\alpha) + h(\alpha) \lvert e^{tr^R} \rvert) \ \text{sign} (e^{tr^R}) [\![ v]\!] \\& + R(\alpha) \lvert [\![ v]\!] \rvert \biggr) \geq 0,
\end{split}
\end{equation}
where the temporal and spatial dependence of the state variables has been dropped to simplify the notation  and the  compliance state function $S(\alpha) = E^{-1}(\alpha)$ has been introduced. The inequality (\ref{eq61}) should be satisfied for all $v$ such that $v(0)=v(L)=0$, all $\beta \geq 0 $ and all $q$ such that $q = q^R + [\![ v ]\!] \delta_{J_v} $.
\begin{itemize}
\item Equilibrium equations \\
Taking $\beta=q^R=0$ and $J(v)=\varnothing$, and integrating by parts the term containing the gradient of $v$ over $ [0,L]\backslash J(u)$, we obtain that the stress is constant over the entire domain:
\begin{equation} \label{eq62}
\text{In} \ [0,L]\backslash J(u): \ \ \sigma'(x)=0,
\end{equation} 
\begin{equation} \label{eq63}
\ \ \ \ \ \ \ \ \text{On} \ J(u): \ \ [\![ \sigma ]\!] (x)=0. 
\end{equation}
\item Transformation yield criterion in the regular part of the domain \\
Taking $v=\beta=0$, hence $[\![ v ]\!]=0$, gives
\begin{equation} \label{eq64}
\begin{split}
    &\int_{[0,L]\backslash J(u)} \biggl(  -\sigma q^R \\&+ (\tau_M(\alpha) + h(\alpha) \lvert e^{tr^R} \rvert) \ \text{sign} (e^{tr^R}) q^R \\& + R(\alpha) \lvert q^R \rvert \biggr) dx \geq 0, \ \ \ \forall q \ \text{smooth}.
    \end{split}
\end{equation}
With the definition (\ref{eq12}) of the transformation stress $X$ we obtain 
\begin{equation} \label{eq65}
\lvert X(\sigma, e^{tr^R}, \alpha) \rvert  - R(\alpha) \leq 0, \ \ \forall x \in [0,L] \backslash J(u).
\end{equation}
\item Transformation yield criterion in the singular part of the domain \\
Taking $\beta = q^R =0$ and $J(v)\neq \varnothing$ and exploiting the equilibrium equations (\ref{eq62})-(\ref{eq63}) leads to 
\begin{equation} \label{eq66}
\begin{split}
&\sum_{J(v)} \biggl( -\sigma [\![ v]\!] + (\tau_M(\alpha) + h(\alpha) \lvert e^{tr^R} \rvert) \ \text{sign} (e^{tr^R}) [\![ v]\!] \\& + R(\alpha) \lvert [\![ v]\!] \rvert \biggr) \geq 0
\end{split}
\end{equation}
or, with (\ref{eq12}),
\begin{equation} \label{eq67}
\sum_{J(v)} \biggl( -X [\![ v]\!] + R(\alpha) \lvert [\![ v]\!] \rvert \biggr) \geq 0.
\end{equation}
Since $J(v)$ and $[\![ v]\!]$ can be chosen arbitrarily, we obtain that the transformation yield criterion must hold everywhere in $[0,L]$.
\item Damage yield criteria \\
Taking $v = q =0$ leads to
\begin{equation} \label{eq68}
\begin{split}
& \int_{[0,L]\backslash J(u)}  \biggl[   \biggl( -\frac{1}{2} S'(\alpha) \sigma^2 + \tau_M'(\alpha) \lvert e^{tr^R} \rvert \\&+ \frac{1}{2} h'(\alpha) \lvert e^{tr^R} \rvert^2 + R'(\alpha) \overline{e^{tr^R}} + w'(\alpha) \biggr) \beta \\& +2w_1 l^2 \alpha' \beta'  \biggr] dx \\
& + \sum_{J(u)} \biggl[ \biggl( \tau_M'(\alpha) \lvert [\![ u]\!] \rvert + \frac{1}{2} h'(\alpha) \lvert [\![ u]\!] \rvert^2 \\&+ R'(\alpha) \overline {[\![ u ]\!]} \biggr) \beta \biggr] \geq 0, \ \ \ \ \forall \beta \geq 0.
\end{split}
\end{equation}
Integrating by parts the term in $\alpha'\beta'$ leads to the damage yield criteria in the regular domain, the singular domain and at the boundaries of the bar:
\begin{equation} \label{eq69}
\begin{split}
&\text{In} \ [0,L]\backslash J(u):\\&
\frac{1}{2}S'(\alpha)\sigma^2 - \tau_M'(\alpha) \lvert e^{tr^R} \rvert  - \frac{1}{2} h'(\alpha) \lvert e^{tr^R} \rvert^2 \\&- R'(\alpha) \overline {e^{tr^R}} - w'(\alpha) + 2 w_1 l^2 \alpha'' \leq 0,
\end{split}
\end{equation} 
\begin{equation} \label{eq70}
\begin{split}
& \text{On} \ J(u): \\&
2 w_1 l^2 [\![ \alpha']\!] - \tau_M'(\alpha) \lvert [\![ u]\!] \rvert - \frac{1}{2} h'(\alpha) \lvert [\![ u]\!] \rvert^2 \\&- R'(\alpha)\overline {[\![ u ]\!]}  \leq 0,
\end{split}
\end{equation} 
\begin{equation} \label{eq71}
\alpha'(0) \leq 0, \ \ \ \ \alpha'(L) \geq 0.
\end{equation} 
\end{itemize}
\subsection{The energy balance}
Assuming that the evolution $t\mapsto \xi_t$ is smooth,  $t\mapsto \overline {e^{tr}}_t$ is obtained from  $t\mapsto e^{tr}_t$ by
\begin{equation} \label{eq72}
\dot {\overline{e^{tr^R}}}_t(x)= \lvert  {\dot{e}^{tr^R}_t} (x)\rvert \ \ \ \forall x \in [0,L]\backslash J(u_t),
\end{equation}
\begin{equation} \label{eq73}
\dot {\overline {[\![ u ]\!]}}_t(x)= \lvert  {[\![ \dot{u}_t]\!]} (x)\rvert \ \ \ \forall x \in J(u_t).
\end{equation}
Expanding the time derivative of the energy and exploiting the equilibrium equation, the energy balance (\ref{eq32}) leads to
\begin{equation} \label{eq74}
\begin{split}
&\int_{[0,L]\backslash J(u)} \biggl[  \sigma (\dot u' - \dot{e}^{tr^R}) \\& + (\tau_M(\alpha) + h(\alpha) \lvert e^{tr^R} \rvert ) \text{sign} (e^{tr^R}) \dot{e}^{tr^R} + R(\alpha) \lvert \dot{e}^{tr^R} \rvert  \\
&+\biggl(   -\frac{1}{2} S'(\alpha) \sigma^2 + \tau_M'(\alpha) \lvert e^{tr^R} \rvert + \frac{1}{2} h'(\alpha) \lvert e^{tr^R} \rvert^2 \\
&+ R'(\alpha) \overline{e^{tr^R}} + w'(\alpha) \biggr) \dot \alpha  +2w_1 l^2 \alpha' \dot\alpha'  \biggr] dx \\
&+ \sum_{J(u)} \biggl[ (\tau_M(\alpha) + h(\alpha) \lvert [\![ u]\!] \rvert ) \text{sign} ([\![ u]\!]) {[\![ \dot u]\!]} + R(\alpha) \lvert {[\![\dot u]\!]} \rvert \\
&+\biggl( \tau_M'(\alpha) \lvert [\![ u]\!] \rvert + \frac{1}{2} h'(\alpha) \lvert [\![ u]\!] \rvert^2 + R'(\alpha) \overline {[\![ u ]\!]} \biggr) \dot \alpha \biggr] \\& - \sigma \dot U = 0.
\end{split}
\end{equation}
Notice that the time and spatial dependence has been removed to simplify the notation. Integrating by parts the term in $\alpha' \dot \alpha'$ and exploiting the identity $\int_{[0,L]\backslash J(u)} \dot u' dx + \sum_{J(u)} {[\![ \dot u]\!]} = \dot U $ gives
\begin{equation} \label{eq75}
\begin{split}
&\int_{[0,L]\backslash J(u)} \biggl(  -\sigma \dot{e}^{tr^R} \\
&+ (\tau_M(\alpha) + h(\alpha) \lvert e^{tr^R} \rvert ) \text{sign} (e^{tr^R}) \dot{e}^{tr^R} \\& + R(\alpha) \lvert \dot{e}^{tr^R} \rvert \biggr) dx \\
& + \int_{[0,L]\backslash J(u)} \biggl(   -\frac{1}{2} S'(\alpha) \sigma^2 + \tau_M'(\alpha) \lvert e^{tr^R} \rvert \\& + \frac{1}{2} h'(\alpha) \lvert e^{tr^R} \rvert^2 + R'(\alpha) \overline{e^{tr^R}}  + w'(\alpha) \\& - 2w_1l^2 \alpha'' \biggr) \dot \alpha  dx \\
&+ \sum_{J(u)} \biggl( -\sigma {[\![ \dot u]\!]} +   (\tau_M(\alpha) + h(\alpha) \lvert [\![ u]\!] \rvert ) \text{sign} ([\![ u]\!]) {[\![ \dot u]\!]} \\&+ R(\alpha) \lvert {[\![ \dot u]\!]} \rvert \biggr)  \\
&+ \sum_{J(u)} \biggl( \tau_M'(\alpha) \lvert [\![ u]\!] \rvert + \frac{1}{2} h'(\alpha) \lvert [\![ u]\!] \rvert^2 + R'(\alpha) \overline {[\![ u ]\!]} \\&- 2w_1 l^2 [\![ \alpha']\!] \biggr) \dot \alpha   \\
& + 2w_1 l^2 (\alpha'(L) \dot \alpha (L) - \alpha'(0) \dot \alpha (0) ) = 0.
\end{split}
\end{equation}
From (\ref{eq75}) the consistency equations and the transformation flow rules are obtained:
\begin{equation} \label{eq76}
\begin{split}
&\text{In} \ [0,L]\backslash J(u): \\&\biggl( \frac{1}{2}S'(\alpha)\sigma^2  - \tau_M'(\alpha) \lvert e^{tr^R} \rvert - \frac{1}{2} h'(\alpha) \lvert e^{tr^R} \rvert^2 \\&- R'(\alpha) \overline {e^{tr^R}} - w'(\alpha) + 2w_1 l^2 \alpha'' \biggr) \dot \alpha =0;
\end{split}
\end{equation} 
\begin{equation} \label{eq77}
\begin{split}
&\text{In} \ [0,L]\backslash J(u): \\&\dot {e}^{tr} \left\{ \begin{array}{rcl} \geq 0 & \mbox{if} & X(\sigma, e^{tr^R},\alpha) = R(\alpha) \\ =0 & \mbox{if} & \lvert X(\sigma,e^{tr^R},\alpha) \rvert -R(\alpha) < 0 \\ \leq 0 & \mbox{if} & X(\sigma,e^{tr^R},\alpha) = -R(\alpha) \end{array}\right.;
\end{split}
\end{equation}
\begin{equation} \label{eq78}
\begin{split}
&\text{On} \ J(u): \\& \biggl( 2w_1 l^2 [\![ \alpha' ]\!] - \tau_M'(\alpha) \lvert [\![ u ]\!] \rvert - \frac{1}{2} h'(\alpha) \lvert [\![ u ]\!] \rvert^2 \\&- R'(\alpha) \overline {[\![ u ]\!]}  \biggr)  \dot \alpha = 0;
\end{split}
\end{equation} 
\begin{equation} \label{eq79}
\begin{split}
&\text{On} \ J(u):  \\&  {[\![\dot u]\!]} \left\{ \begin{array}{rcl} \geq 0 & \mbox{if} & X(\sigma, [\![u]\!],\alpha) = R(\alpha) \\ =0 & \mbox{if} & \lvert X(\sigma,[\![u]\!],\alpha) \rvert -R(\alpha) < 0 \\ \leq 0 & \mbox{if} & X(\sigma,[\![u]\!],\alpha) = -R(\alpha) \end{array}\right..
\end{split}
\end{equation}
\begin{equation} \label{eq80}
\alpha'(0)\dot \alpha(0) = 0, \ \ \ \alpha'(L)\dot \alpha(L) = 0.
\end{equation}
\bibliography{sn-bibliography}

\end{document}